\documentclass[aps,pre,amsmath,amssymb,twocolumn,groupedaddress,showpacs,eqsecnum]{revtex4}
\usepackage{graphicx}
\usepackage[utf8]{inputenc}
\usepackage[caption = false]{subfig}
\usepackage{float}
\usepackage{amsmath,amsfonts,amsthm,bm}
\usepackage{braket}

\usepackage{epsf}
\usepackage[toc]{appendix}
\newcommand{\be}{\begin{equation}}
\newcommand{\ee}{\end{equation}}
\newcommand{\ba}{\begin{eqnarray}}
\newcommand{\ea}{\end{eqnarray}}
\newcommand{\baa}{\begin{eqnarray}}
\newcommand{\eaa}{\end{eqnarray}}
\newcommand{\ed}{\end{document}}

\renewcommand{\baselinestretch}{1.2}
\date{\today}

\begin{document}
\title{Fast-forward approach to adiabatic quantum dynamics of regular spin clusters: \\
nature of geometry-dependent driving interactions}
\author{ Iwan Setiawan$^{1,2}$, Bobby Eka Gunara$^{2}$, Sanat Avazbaev$^{3,4}$and Katsuhiro Nakamura$^{5,6}$}
\affiliation{$^{(1)}$Department of Physics Education, University of Bengkulu, Kandang Limun, Bengkulu 38371, Indonesia\\\\
$^{(2)}$Department of Physics, Institut Teknologi Bandung,  Jalan Ganesha, Bandung 40132, Indonesia\\
$^{(3)}$Faculty of Physics and Mathematics, Tashkent State Pedagogical University,  27 Bunyodkor Street, Tashkent 100070, Uzbekistan\\
$^{(4)}$Yeoju Technical Institute in Tashkent, 156 Usmon Nosir Street, Tashkent 100056, Uzbekistan\\
$^{(5)}$Faculty of Physics, National University of Uzbekistan, Vuzgorodok, Tashkent 100174, Uzbekistan\\
$^{(6)}$Department of Applied Physics, Osaka City University, Sumiyoshi-ku, Osaka 558-8585, Japan\\
}

\begin{abstract}
The fast forward scheme of
adiabatic quantum dynamics is applied to finite regular spin clusters with various geometries and the  nature of driving interactions is elucidated.  
The fast forward is the quasi-adiabatic dynamics
guaranteed by regularization terms added to the reference Hamiltonian, followed by a rescaling of time with use of a large scaling factor. 
With help of the regularization terms consisting of
pair-wise and 3-body interactions, we apply the proposed formula (Phys. Rev. A $\bf{96}$, 052106(2017)) to regular triangle and open linear chain for $N=3$ spin systems, and to triangular pyramid, square, primary star graph and open linear  chain for $N=4$ spin systems. 
The geometry-induced symmetry greatly decreases the rank of coefficient matrix of the linear algebraic equation for regularization terms.  Choosing a transverse Ising Hamiltonian as a reference, we find: (1) for $N=3$ spin clusters, the driving interaction consists of only the geometry-dependent pair-wise interactions and there is no need for the 3-body interaction; (2) for $N=4$ spin clusters,  the geometry-dependent pair-wise interactions again constitute major part of the driving interaction, whereas the universal 3-body interaction free from the geometry is necessary but plays a subsidiary role. Our scheme predicts the practical driving interaction in accelerating the adiabatic quantum dynamics of structured regular spin clusters.
\end{abstract}
\maketitle

\section{INTRODUCTION}

Effectively manipulating and optimizing the dynamics of given systems constitutes one of big experimental and theoretical subjects in the current technology.
In particular, it is a challenging theme to find suitable driving fields for tailoring a quantum system to rapidly generate a target state from a given initial state.  In designing quantum computers, the acceleration of adiabatic quantum dynamics is  desirable because the coherence of systems is degraded by their interaction with the environment. Since naive numerical trial-and-error methods are time- and resource-consuming, we must deeply understand relevant quantum dynamics  to find useful schemes for such accelerations.
In this context,  various researches on the way to the shortcut to adiabaticity (STA) have been developed, which include invariant-based inverse engineering \cite{1,2,3}, transitionless counter-diabatic (CD) driving \cite{4,5,6},  fast-forward approach \cite{7,8,9}, and variational methods to generate approximate CD protocols\cite{10,11,12}.

The fast-forward theory proposed by Masuda and Nakamura \cite{7} was originally concerned with acceleration of general reference quantum dynamics. 
This theory was developed to accelerate the adiabatic quantum dynamics by introducing the large time-scaling factor in the quasi-adiabatic dynamics
guaranteed by regularization terms added to the reference Hamiltonian \cite{8,9}, and was then used to enhance the quantum tunneling power \cite{13} and to construct the non-equilibrium equation of state under a rapid piston \cite{14}. The relation between the fast-forward approach and other methods was rigorously investigated in \cite{15}.

Recently, we proposed a fast forward scheme of adiabatic spin dynamics \cite{16}. 
Confining to a single and two spin systems there, we showed the acceleration of  Landau-Zener transition and that of a generation of entangled states, as can be shown in other methods \cite{4,5,6,3extra,10,paul,stef} .

The fast forward scheme of adiabatic quantum dynamics has advantages as addressed by \cite{TK1,TK2}:
(1) No need of writing the driving interaction in the spectral representation with use of full spectral properties of  given spin systems. No necessity of worrying about the divergence of the driving interaction due to the level crossing;
(2) A great flexibility in choosing the regularization Hamiltonian which leads to the driving interaction. Namely, users can specify the regularization Hamiltonian by themselves so as to satisfy the core equation (see Eq.(\ref{sum2}) of this paper).
The latter advantage will play an important role  when we shall investigate spin clusters of various geometries. However, no technical guide was so far presented in solving the core equation for unknown regularization terms.

Within a framework of the transitionless CD driving \cite{4,5,6},  on the other hand, there exist intensive works on a linear chain of many quantum spins described by the Ising model in a transverse field  \cite{ex1,ex2,ex3,ex33} and the related model \cite{campb}, which showed the complicated non-local multi-body CD terms that are hard to achieve in experiment. While a variational method to generate approximate local CD protocols\cite{11,12} is being cultivated, it is timely to 
sharpen the fast-forward approach by showing a guiding principle to manage spin clusters with various geometries on the basis of the proposed formula in \cite{16}.

In this paper the fast forward scheme of adiabatic dynamics is applied to regular spin clusters of various geometries with number of spins $N$ up to 4, i.e., regular triangle and open linear chain for $N=3$ spins, and triangular pyramid, square, primary star graph and open linear  chain for $N=4$ spins. (Note: the geometry is irrelevant for systems with $N=1$ and $2$ spins.)  Choosing the Hamiltonian for a transverse Ising model as a reference,  we shall reveal the nature of driving interactions. In Section \ref{FFADspin}, a brief summary is given on the fast forward scheme of adiabatic quantum spin dynamics. In Section \ref{Explicit results} we propose a candidate regularization Hamiltonian consisting of geometry-dependent pair-wise interactions and a universal 3-body interaction, and describe a method of solving the linear algebraic equation for regularization terms. Sections \ref{Three spins} and \ref{Four spins} are devoted to the analysis of spin clusters of various geometries with $N=3$ and $N=4$, respectively. Summary and discussions are given in Section \ref{concl}.  Appendix  \ref{apdB} gives matrices for some regularization Hamiltonians.

\section{Fast-forward scheme of adiabatic spin dynamics}\label{FFADspin}

For self-containedness, we shall sketch the fast forward scheme of adiabatic spin dynamics\cite{16}.
Our strategy is as follows: (i) A given original (reference) Hamiltonian  $H_{0}$ is assumed to 
change adiabatically and to generate a stationary state $\Psi_{0}$, which is an eigenstate
of the time-independent Schr\"{o}dinger equation with the instantaneous Hamiltonian.
Then $H_{0}$ is regularized so that $\Psi_{0}$ should satisfy the time-dependent Schr\"{o}dinger equation (TDSE); (ii) Taking $\Psi_{0}$ as a reference state, we shall rescale time in TDSE with use of the scaling factor $\alpha\left(t\right)$, where the mean value $\bar{\alpha}$ of the infinitely-large time scaling factor $\alpha(t)$ will be chosen to compensate the infinitesimally-small growth rate $\epsilon$ of the quasi-adiabatic parameter and to satisfy $\bar{\alpha} \times \epsilon=finite$.

Consider the Hamiltonian for spin systems to be characterized by a slowly time-changing parameter $R(t)$ such as the exchange interaction, magnetic field, etc. Then we can study the eigenvalue problem for the time-independent Schr\"{o}dinger equation :
\begin{equation}\label{satu}
H_0(R)\textbf{C}^{(n)}(R)=E_n(R)\textbf{C}^{(n)}(R)
\end{equation}
with
\begin{equation}\label{satu-1}
\textbf{C}^{(n)}(R)=
\begin{pmatrix} 
  C^{(n)}_1(R) \\  \vdots\\ 
  C^{(n)}_N(R)
\end{pmatrix}, 
\end{equation}
where 
\begin{equation}
R\equiv R(t)= R_0 + \epsilon t  
\end{equation}
is the adiabatically-changing parameter with $\epsilon \ll 1$. In Eq.(\ref{satu}), 
$n$ stands for the quantum number for each eigenvalue and eigenstate.
Let us assume 
\begin{equation}\label{psi0}
\Psi^{(n)}_0(R(t)) =\textbf{C}^{(n)}(R(t))e^{-\frac{i}{\hbar}\int_{0}^{t}E_n(R(t'))dt'} e^{i\xi_n(R(t))},
\end{equation}
to be a quasi-adiabatic state, i.e., adiabatically evolving state, where $\xi_n$ is the adiabatic phase:
\begin{eqnarray}\label{adiabatic}
\xi_n(R(t))= i \int_{0}^{t} dt' \textbf{C}^{(n)\dagger}\partial_t\textbf{C}^{(n)} = i \epsilon \int_{0}^{t} dt'  \textbf{C}^{(n)\dagger}\partial_R\textbf{C}^{(n)}. \nonumber\\ 
\end{eqnarray}

$\Psi^{(n)}_0(R(t))$ in Eq.(\ref{psi0}) is not a solution of TDSE.
To make  it  to satisfy the TDSE, 
we must regularize the Hamiltonian as
\begin{equation}\label{reg}
H_0^{reg}(R(t)) = H_0(R(t)) + \epsilon\mathcal{\tilde{H}}_n(R(t)).
 \end{equation}
Then TDSE becomes
\begin{equation}\label{adia-TDSE}
i \hbar \frac{\partial}{\partial t} \Psi^{(n)}_0(R(t))=(H_0+\epsilon \mathcal{\tilde{H}}_n)\Psi^{(n)}_0(R(t)).
\end{equation}
Here  $\mathcal{\tilde{H}}_n$ is the $n$-th state-dependent regularization term.
Substituting $\Psi^{(n)}_0(R(t))$ in Eq.(\ref{psi0}) into the above TDSE, we see the eigenvalue problem in Eq.(\ref{satu}) in  order of $O(\epsilon^0)$,
and the algebraic equation for $\mathcal{\tilde{H}}_n$,
\begin{equation}\label{sum2}
\mathcal{\tilde{H}}_n\textbf{C}^{(n)}(R)= 
i \hbar \partial_R\textbf{C}^{(n)}(R) -i\hbar ( \textbf{C}^{(n)\dagger}\partial_R\textbf{C}^{(n)})
\textbf{C}^{(n)}(R),
\end{equation}
in order of $O(\epsilon^1)$. Equation (\ref{sum2}) is the core  of the present study.
The state in Eq.(\ref{psi0}) and TDSE in Eq.(\ref{adia-TDSE}) are working on a very slow time scale.
We shall innovate them so that they can work on a laboratory time scale.
 
With time $t$ rescaled by the advanced time $\Lambda(t)$, the fast-forward state is introduced as
\begin{eqnarray}\label{psiff}
\Psi^{(n)}_{FF}(t) &\equiv& \Psi^{(n)}_0(R(\Lambda(t))) \nonumber\\
&=& \textbf{C}^{(n)}(R(\Lambda(t)))
 e^{-\frac{i}{\hbar}\int_{0}^{t}E_n(R(\Lambda(t')))dt'} e^{i\xi_n(R(\Lambda(t)))}, \nonumber\\
\end{eqnarray}
where $\Lambda(t)$ is defined by
\begin{equation}\label{lamda}
\Lambda(t) =  \int_{0}^{t} \alpha (t')dt',
\end{equation}
with the standard time $t$. $\alpha(t)$ is an arbitrary magnification time-scale factor which satisfies $\alpha(0)$ = 1, $\alpha(t) > 1(0 < t < T_{FF})$ and $\alpha(t)$ = $1 (t\geq T_{FF})$. 
For a long final time $T$ in the original adiabatic dynamics, we can consider the fast forward dynamics with a new time variable which reproduces the target state $\Psi^{(n)}_0(R(T))$ in a shorter final time $T_{FF}$ defined by
\begin{equation}
T = \int_{0}^{T_{FF}}\alpha(t)dt.
\end{equation}
The simplest expression for $\alpha(t)$ in the fast-forward range ($0\leq t \leq T_{FF} $) is  given by \cite{8} as :
\begin{equation}
\alpha(t) = \bar{\alpha}-(\bar{\alpha}-1) \cos(\frac{2 \pi}{T_{FF}}t),
\end{equation}
where $\bar{\alpha}$ is the mean value of $\alpha(t)$ and is given by $\bar{\alpha} = T/T_{FF}$. 

Then by taking the time derivative of $\Psi^{(n)}_{FF}$ in Eq.(\ref{psiff})
and using the equalities $\partial_t \textbf{C}^{(n)}(R(\Lambda(t))) = 
\alpha \epsilon \partial_R \textbf{C}^{(n)}$ and $\partial_t\xi_n(R(\Lambda(t)))$ = $i \textbf{C}^{(n)\dagger}\partial_t\textbf{C}^{(n)}$ = $i \alpha \epsilon \textbf{C}^{(n)\dagger} \partial_R \textbf{C}^{(n)}$, we have
\begin{eqnarray}
i \hbar \dot{\Psi}^{(n)}_{FF} &=& \Big[ i \hbar \alpha \epsilon \left(\partial_R\textbf{C}^{(n)}
-(\textbf{C}^{(n)\dagger}\partial_R\textbf{C}^{(n)})\textbf{C}^{(n)}\right)+ E \textbf{C}^{(n)}\Big]\nonumber\\
&\times& e^{-\frac{i}{\hbar}\int_{0}^{t}E_n(R(\Lambda(t')))dt'} e^{i\xi_n(R(\Lambda(t)))}.
\end{eqnarray}
The first and second terms in the angular bracket on the r.h.s are replaced by $\alpha\epsilon\mathcal{\tilde{H}}_n\textbf{C}^{(n)}(R(\Lambda(t)))$ and $H_0\textbf{C}^{(n)}(R(\Lambda(t)))$, respectively, by using Eqs.(\ref{sum2}) and (\ref{satu}). Using the definition of $\Psi^{(n)}_{FF}(t)$ and taking the asymptotic limit
$\bar{\alpha} \rightarrow \infty$ and $\epsilon \rightarrow 0$ under the constraint $\bar{\alpha} \cdot \epsilon \equiv \bar{v}= finite$, we obtain 
\begin{eqnarray}\label{TDSE}
i \hbar \frac{\partial \Psi^{(n)}_{FF}}{\partial t} &=& \left(H_0(R(\Lambda(t)))
+v(t) \mathcal{\tilde{H}}_n(R(\Lambda(t))) \right) \Psi^{(n)}_{FF}\nonumber\\
&\equiv&  H^{(n)}_{FF} \Psi^{(n)}_{FF}. 
\end{eqnarray}
Here $v(t)$ is a velocity function available from $\alpha(t)$ in the asymptotic limit:
\begin{eqnarray}\label{vt}
v(t) = \lim_{\epsilon\to 0, \bar{\alpha} \to \infty} \epsilon \alpha(t) = \bar{v}\left(1-\cos \frac{2 \pi}{T_{FF}}t\right).
\end{eqnarray}
Consequently, for $0  \le t \le T_{FF}$,
\begin{eqnarray}\label{lambda2}
R(\Lambda(t))&=&R_0+\lim_{\epsilon\rightarrow 0, \bar{\alpha}\rightarrow \infty}\varepsilon\Lambda(t) 
=R_{0}+\int^{t}_{0}v(t')dt'\nonumber\\
&=& R_{0}+\bar{v}\left[t-\frac{T_{FF}}{2\pi}\sin\left(\frac{2\pi}{T_{FF}}t\right)\right].
\end{eqnarray}
$H^{(n)}_{FF}$ is the fast-forward Hamiltonian and $\mathcal{\tilde{H}}_n$ is the regularization term obtained from Eq.(\ref{sum2}) to generate the fast-forward scheme in spin system. 
Eqs. (\ref{psiff}) and (\ref{TDSE}) work on a laboratory time scale.

There is a relationship between our formula for $\mathcal{\tilde{H}}_n$ in Eq.(\ref{sum2}) and Demirplak-Rice-Berry (DRB)'s formula \cite{4,5,6} for the CD term $\mathcal{H}$.
If there is a $n$-independent regularization term $\mathcal{\tilde{H}}$ among $\{\mathcal{\tilde{H}}_n\}$, we can define $\mathcal{H} \equiv v(t) \mathcal{\tilde{H}}$ 
with use of $v(t) = \frac{\partial R(\Lambda(t))}{\partial t}$. Then Eq.(\ref{sum2}) gives a solution $\mathcal{H} $ which agrees with DRB's formula  for the CD term (See the proof in \cite{16}). 
It should be noted, however, that the above correspondence works well only in the case that we can find $n$-independent regulariztion terms $\mathcal{\tilde{H}}$ among 
$\{\mathcal{\tilde{H}}_n\}$. 
Using the above notion, one may call $v(t) \mathcal{\tilde{H}}_n$ as a state-dependent CD term.
Hereafter we shall be concerned with the fast forward of adiabatic dynamics of one of the adiabatic states (i.e., the ground state) and therefore the suffix $n$ in $ \mathcal{\tilde{H}}_n$ will be suppressed.

\begin{figure}[!h]
\subfloat[]{%
  \includegraphics[width=1.8in]{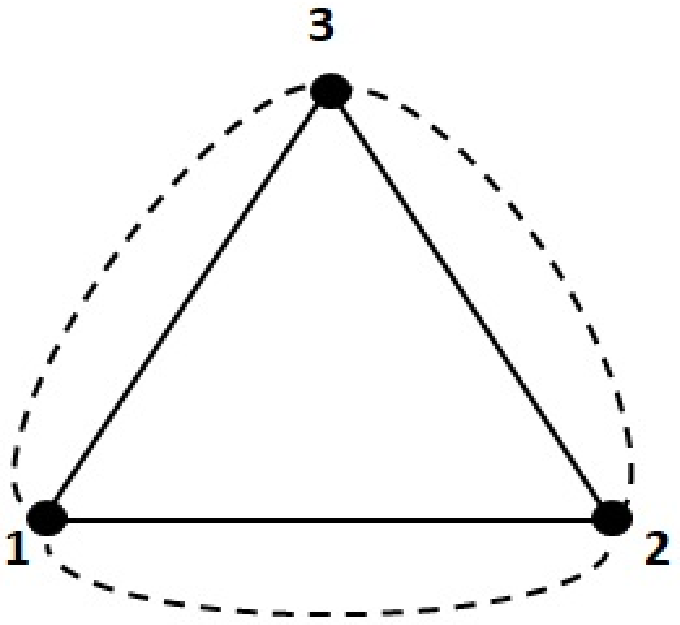}
}
\hfill
\subfloat[]{%
 \includegraphics[width=2in]{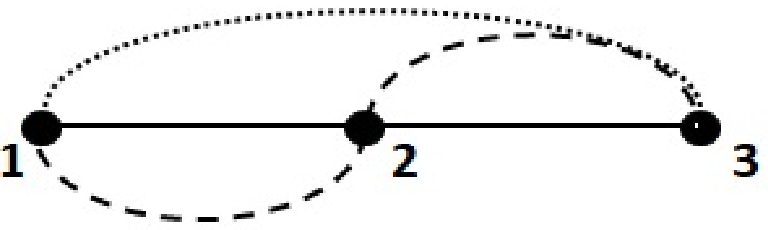}%
}
\caption{(a) Regular triangle; (b) Open linear 3 spin chain. Solid lines stand for the original exchange interactions. Dashed and dotted lines mean the pair-wise regularization interactions. Each line species denotes the geometrically-identical regularization interactions.} \label{3spinmodel}
\end{figure}

\begin{figure}[!h]
\subfloat[]{%
  \includegraphics[width=1.8in]{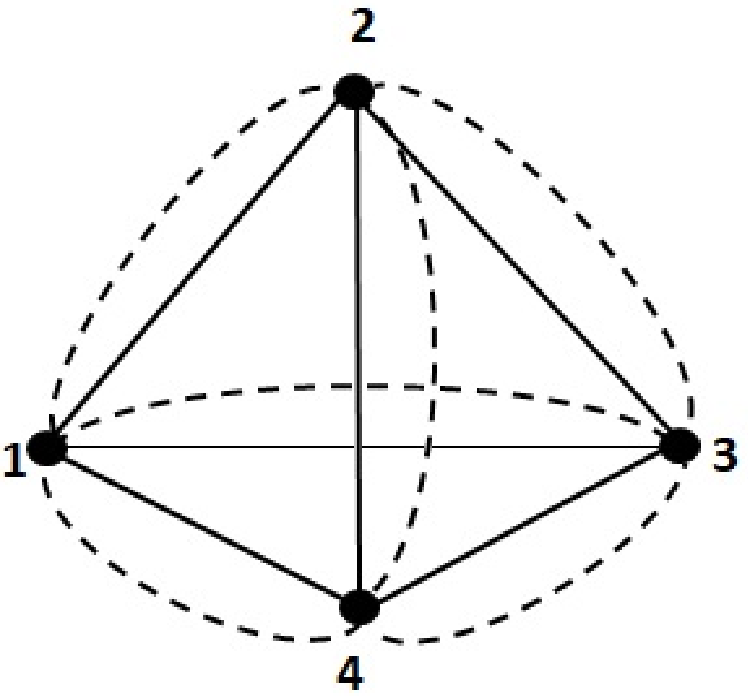}
}
\hfill
\subfloat[]{%
 \includegraphics[width=2in]{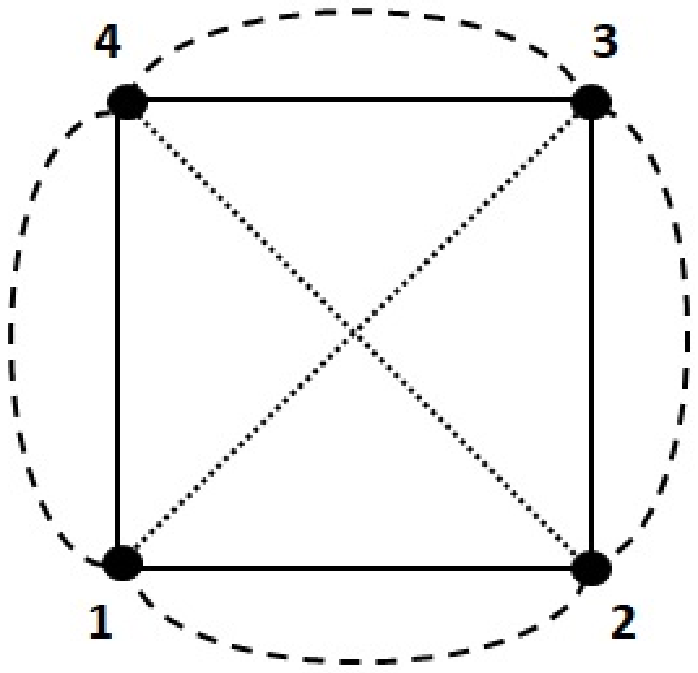}%
}
\hfill
\subfloat[]{%
 \includegraphics[width=1.8in]{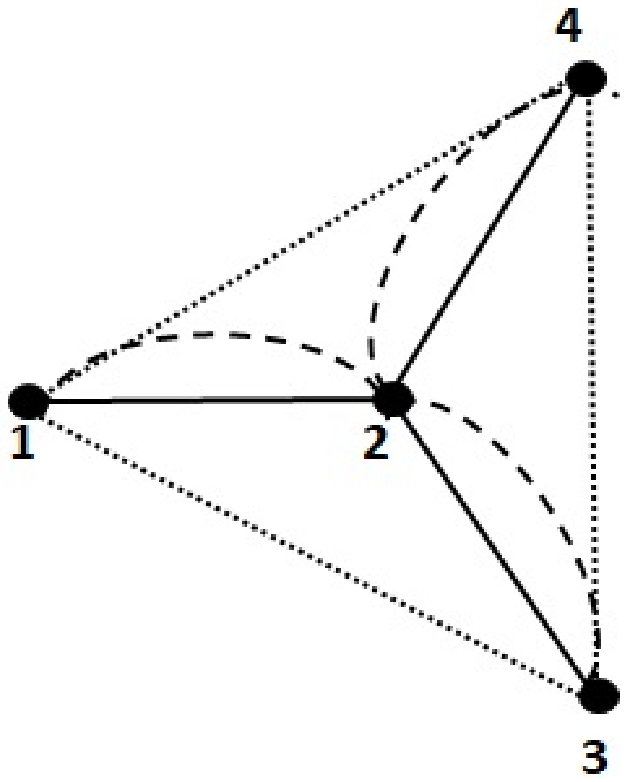}%
}
\hfill
\subfloat[]{%
 \includegraphics[width=2in]{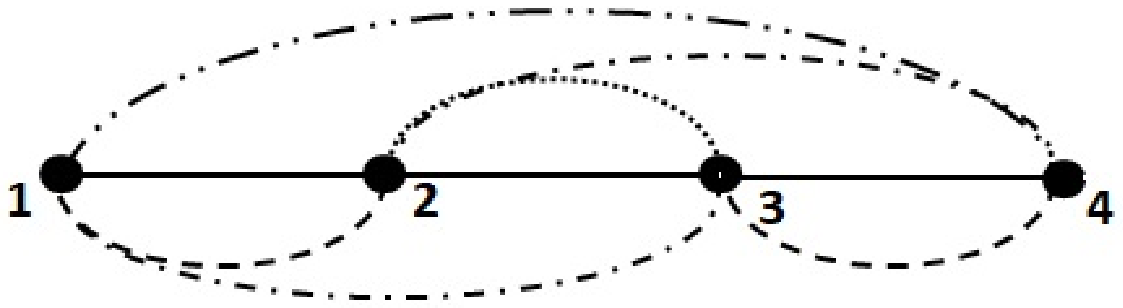}%
}
\caption{(a) Triangular pyramid; (b) Square; (c) Primary star graph; (d) Open linear 4 spin chain. Solid lines stand for the original exchange interactions. Dashed, dotted, dotted dashed and double-dotted dashed lines mean the pair-wise regularization interactions. Each line species denotes the geometrically-identical regularization interactions.} \label{4spinmodel} 
\end{figure}

\section{Fast-forward driving interactions for spin clusters of various geometries}\label{Explicit results}
To begin with, let us explain the method of solving the linear algebraic equation for unknown regularization 
terms in Eq. (\ref{sum2}).
Then in the succeeding Sections, we shall treat regular spin clusters of various geometries with $N$ up to 4, i.e., regular triangle and open linear chain for $N=3$ spins (see Fig.\ref{3spinmodel}), and triangular pyramid, square, primary star graph and open linear  chain for $N=4$ spins (see Fig.\ref{4spinmodel}). 
Our scheme is free from obtaining all eigenvectors for a given adiabatic Hamiltonian.
As shown in the core equation in Eq. (\ref{sum2}), we need only information of  a single eigenstate, typically of the ground state. 

As an original (reference) model, we choose the transverse Ising mode,  whose Hamiltonian for $N$ spin systems is written as
\begin{equation}\label{SIM}
H_0 = J(R(t)) \sum_{(i,j) \in N.N.}\sigma_i^z \sigma_j^z- \frac{1}{2}B_x(R(t))\sum_{i=1}^{N}\sigma_i^x,
\end{equation}
where  $J(R(t))=R(t)=R_0+\epsilon t$ and $B_x(R(t))=B_0-R(t)$ with $\epsilon \ll 1$ are adiabatically-changing exchange interaction and transverse magnetic field, respectively.
$(i,j) \in N.N.$ means nearest-neighbouring pairs.
Using the spin configuration bases, the dimension of Hilbert space is $2^N$. 

Energy matrix corresponding to the Hamiltonian in Eq.(\ref{SIM}) is real symmetric, which makes the eigenstates real, and the ground state is expressed by 
the real components $\{C_k: k=1, \cdots, 2^N\}$. 
This, in combination with the fact that the length of the corresponding eigenvector is constant and equal to 1, leads to the conclusion that the adiabatic phase $\xi_n$ in Eq.(\ref{adiabatic}) is zero in all spin clusters in the present work. 
Further, because of the geometrical symmetry of spin clusters in Figs. \ref{3spinmodel} and \ref{4spinmodel}, some of the components $C_k$s are degenerate which reduce the number of independent equations in the core equation in Eq. (\ref{sum2}).

As for  the unknown regularization term ($\mathcal{\tilde{H}}$) in Eq.(\ref{sum2}) , we must impose
a form which makes its matrix elements pure imaginary because the right-hand side of
Eq.(\ref{sum2}) is now pure imaginary.  Among several possibilities, we assume the regularization term consisting of  pair-wise interactions described by $\tilde{W}_{ij}^{yz} = \tilde{W}_{ij}^{yz}(\epsilon t)$ and 3-body interactions $\tilde{Q}_{ijk}^{xyz} = \tilde{Q}_{ijk}^{xyz}(\epsilon t)$.  Other possible contributions such as
a single-particle energy due to $y$-component of the magnetic field ($\tilde{B}_y$), pair-wise interaction $\tilde{W}_{ij}^{xy}$  and 3-body interaction $\tilde{Q}_{ijk}^{xxy}$
lead to incompatible algebraic equations in Eq.(\ref{sum2}), and should be excluded.
The candidate for regularization Hamiltonian $\mathcal{\tilde{H}}$ then takes the following form :
\begin{eqnarray}\label{cdterm}
\mathcal{\tilde{H}} &=& \sum_{(i,j) \in all}\tilde{W}_{ij}^{yz} (\sigma_i^y \sigma_j^z + \sigma_i^z \sigma_j^y) \nonumber\\
&+& \sum_{(i,j,k) \in all}\tilde{Q}_{ijk}^{xyz} (\sigma_i^x \sigma_j^y + \sigma_i^y \sigma_j^x)\cdot \sigma_k^z,
\end{eqnarray}
where $(i,j) \in all$ and $(i,j,k) \in all$ mean all possible combinations (not permutations), and are not limited to nearest neighbours.  
The  3-body interaction here is not brought as a result of the truncation of long-range and multi-body counter-diabatic interactions, but is introduced in advance to make the core equation solvable.

Since regular spin clusters have geometric symmetry, some of the interactions ($\tilde{W}_{ij}^{yz}$) are degenerate as shown in Figs. \ref{3spinmodel} and \ref{4spinmodel}, and the reduced number of independent interactions should be equal to the number of independent equations in Eq.(\ref{sum2}). In the present paper, the 3-body interaction will play a subsidiary role.  Below we shall solve the regularization terms and obtain the fast-forward Hamiltonian for spin clusters of various geometries.
 
\section{Regular triangle and open linear 3 spins}\label{Three spins}
In this Section we investigate a regular triangle and open linear 3 spins in Fig. \ref{3spinmodel}.
We use the spin configuration bases as $\Ket{1}=\Ket{\uparrow\uparrow\uparrow} $, $\Ket{2}=\Ket{\uparrow\uparrow\downarrow} $, $\Ket{3}=\Ket{\uparrow\downarrow\uparrow}$, $\Ket{4}=\Ket{\downarrow\uparrow\uparrow}$, $\Ket{5}=\Ket{\uparrow\downarrow\downarrow}$, $\Ket{6}=\Ket{\downarrow\uparrow\downarrow}$, $\Ket{7}=\Ket{\downarrow\downarrow\uparrow}$  and $\Ket{8}=\Ket{\downarrow\downarrow\downarrow}$.  

\subsection{Regular triangle}
\begin{figure}[!h]
\subfloat[]{%
\includegraphics[width=2.5in]{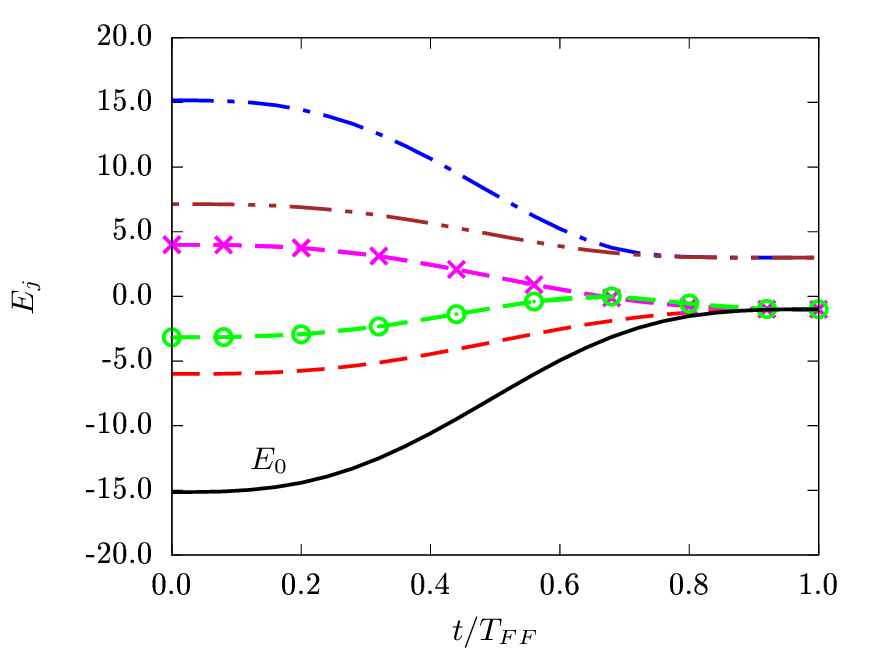}%
 }
\hfill
\subfloat[]{%
\includegraphics[width=2.5in]{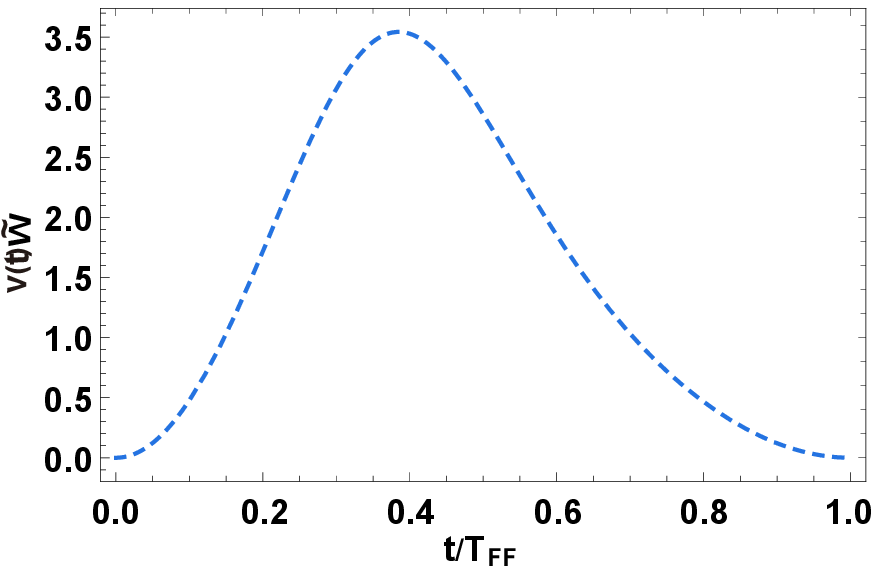} %
 }
\hfill
\subfloat[]{%
\includegraphics[width=2.5in]{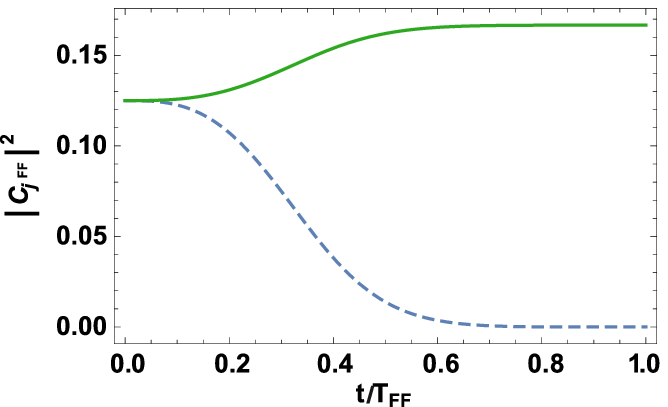}%
}
\caption{
The time dependence in the case of the regular triangle in the fast-forward time range where we choose $J=R(\Lambda(t))$ and $B_x = B_0-R(\Lambda(t))$ with 
$R(\Lambda(t))$ defined in Eq.(\ref{lambda2}). $B_0 =10$ and $\bar{v}= 100$.
 $T_{FF}=0.1$ and $R_0=0$.
(a) All eight eigenvalues. From the bottom, the 2nd and 4th lines are each doubly degenerate;
(b) Regularization term $v(t)\tilde{W}$; 
(c) Probability amplitudes for the solution $\Psi_{FF}(t)$ of TDSE, $|C_2^{FF}|^2=|C_3^{FF}|^2=|C_4^{FF}|^2=|C_5^{FF}|^2=|C_6^{FF}|^2=|C_7^{FF}|^2$ (solid line) 
and $ |C_1^{FF}|^2=|C_8^{FF}|^2$ (dashed line).
} \label{fig-triangl}
\end{figure}

In the case of the regular triangle, the eigenvalue for the ground state is $E_0 = -\sqrt{B_x^2+2 B_x J+4 J^2}-\frac{B_x}{2}+J$.
We have confirmed in Fig. \ref{fig-triangl}(a) that all eight eigenvalues show no mutual energy crossing in the fast-forward time range where we choose $J(R(\Lambda(t)))\equiv R(\Lambda(t))$
and $B_x(R(\Lambda(t)))\equiv B_0 - R(\Lambda(t))$ with $R(\Lambda(t))$ defined in Eq.(\ref{lambda2}). 

The components of the eigenvector for the ground state are  :\\
$ C_1 = V_1 \zeta$, $C_2 =V_2 \zeta$, $C_3 = V_3 \zeta$, $C_4 = V_4 \zeta$, $C_5 = V_5 \zeta$, $C_6 =V_6 \zeta$, $C_7 = V_7 \zeta$, $C_8 = V_8 \zeta $, where $V_1 = V_8 = 1$, $
V_2 = V_3 = V_4 =V_5 = V_6 =V_7 = \frac{2\sqrt{B_x^2+2 B_x J+4 J^2}+B_x +4J}{3 B_x}$, and  
$\zeta=\frac{1}{\sqrt{2 +6 V_2^2}}$. 

Here we see the symmetry: $C_1=C_8$, $C_2 =C_3=C_4=C_5=C_6=C_7$. From $R$-derivative of the normalization ($\sum_{j=1}^8C_j^2=2C_1^2+6C_2^2 = 1$), we see
\begin{equation}\label{adi}
C_1 \frac{\partial C_1}{\partial R}+ 3 C_2\frac{\partial C_2}{\partial R} =0 ,
\end{equation}
and then the adiabatic phase $\xi = 0$. 

As for  the regularization Hamiltonian for the regular triangle, we can proceed without having recourse to the 3-body interaction. Three $\tilde{W}_{ij}^{yz}$s should be identical due to the triangular symmetry in Fig. \ref{3spinmodel}(a). Therefore the unknown pairwise interaction is only one: $\tilde{W}\equiv \tilde{W}_{ij}^{yz}$, independent of the pairs $(i,j)$.

By using the spin configuration bases as above,  the regularization Hamiltonian in Eq.(\ref{cdterm}) is characterized by
the matrix elements: $\mathcal{\tilde{H}}_{1j}=-\mathcal{\tilde{H}}_{j1}=-2i\tilde{W}$ with $j=2,3,4$, 
$\mathcal{\tilde{H}}_{8j}=-\mathcal{\tilde{H}}_{j8}=-2i\tilde{W}$ with $j=5,6,7$ and all other elements $=0$.
The explicit expression for $\mathcal{\tilde{H}}$ will help us to solve Eq.(\ref{sum2}). 

Due to the symmetry of $\{C_j\}$, the number of independent equations are only two in Eq.(\ref{sum2}) :
\begin{eqnarray}\label{aku2x2}
-6\tilde{W}C_2&=&\hbar \frac{\partial  C_1}{\partial R}, \nonumber\\
2\tilde{W}C_1 &=&\hbar \frac{\partial  C_2}{\partial R}. 
\end{eqnarray}
Noting the normalization-assisted relation in Eq.(\ref{adi}), one of the above two equations becomes trivial, and Eq.(\ref{aku2x2}) has the solution:
\begin{eqnarray}\label{W2}
\tilde{W}&=&\hbar \frac{\partial_R C_2}{2C_1}=\hbar(C_1 \partial_RC_2 - C_2 \partial_RC_1) \nonumber \\
&=&\frac{B_x \frac{\partial J}{\partial R}-J\frac{\partial B_x}{\partial R}}{4 (B_x^2+2 B_x J+4 J^2)}.         
\end{eqnarray}
The second equality above is due to the normalization condition and Eq.(\ref{adi}).
Including the regularization term followed by rescaling of time, the fast forward Hamiltonian is written as 
\begin{equation}\label{HFF0}
H_{FF}= H_0(R(\Lambda(t)))+v(t) \tilde{\mathcal{H}} (R(\Lambda(t)))
\end{equation}
with $H_0$ = $J(R(\Lambda(t))) (\sigma_1^z \sigma_2^z+ \sigma_2^z \sigma_3^z+ \sigma_3^z \sigma_1^z)- \frac{1}{2}(\sigma_1^x+ \sigma_2^x+\sigma_3^x)B_x(R(\Lambda(t)))$, and $v\mathcal{\tilde{H}}$ = $v(t) \tilde{W}(R(\Lambda(t)))\big[ (\sigma_1^y \sigma_2^z+ \sigma_1^z \sigma_2^y)+(\sigma_2^y \sigma_3^z+ \sigma_2^z \sigma_3^y)+(\sigma_3^y \sigma_1^z+ \sigma_3^z \sigma_1^y)\big]$.

The fast forward Hamiltonian guarantees the fast forward of the adiabatic dynamics of the ground state wave function.
Figures \ref{fig-triangl}(b) and \ref{fig-triangl}(c)  show the time dependence of the  regularization term and that of the wave function, respectively. The wave function starts from the ground state with $J=0$, i.e., $C_1 = C_2 = C_3 = C_4 = C_5= C_6= C_7= C_8= \frac{1}{2\sqrt{2}}$.
The initial state  is a linear combination of $\Ket{\uparrow\uparrow\uparrow} $, $\Ket{\uparrow\uparrow\downarrow} $, $\Ket{\uparrow\downarrow\uparrow}$, $\Ket{\downarrow\uparrow\uparrow}$, $\Ket{\uparrow\downarrow\downarrow}$, $\Ket{\downarrow\uparrow\downarrow}$, $\Ket{\downarrow\downarrow\uparrow}$  and $\Ket{\downarrow\downarrow\downarrow}$ states.  
 As $J$  is increased from $0$ and $B_x$ is decreased, the system rapidly changes to the final state, a linear combination of reduced bases  $\Ket{\uparrow\uparrow\downarrow} $, $\Ket{\uparrow\downarrow\uparrow}$, $\Ket{\downarrow\uparrow\uparrow}$, $\Ket{\uparrow\downarrow\downarrow}$, $\Ket{\downarrow\uparrow\downarrow}$, and $\Ket{\downarrow\downarrow\uparrow}$.  In Fig. \ref{fig-triangl} (c)  the solution $\Psi_{FF}(t)$ of TDSE in Eq.(\ref{TDSE})  has reproduced the time-rescaled ground state wave function, which means the perfect fidelity of $\Psi_{FF}(t)$ during the fast-forward time range $0 \leq t \leq T_{FF}$.

\subsection{Open linear 3 spin chain}
\begin{figure}[!h]
\subfloat[]{%
  \includegraphics[width=2.5in]{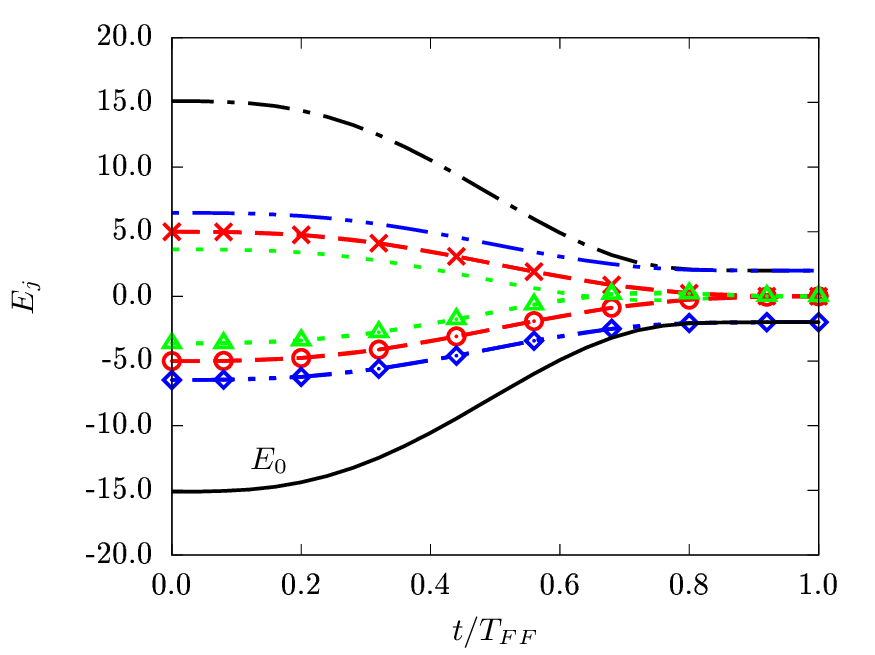}
}
\hfill
\subfloat[]{%
  \includegraphics[width=2.5in]{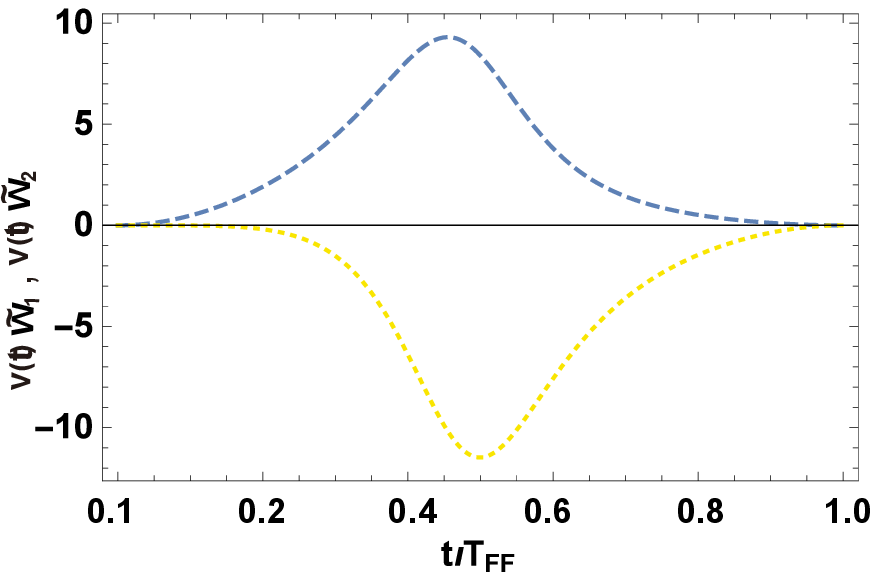}
}
\hfill
\subfloat[]{%
  \includegraphics[width=2.5in]{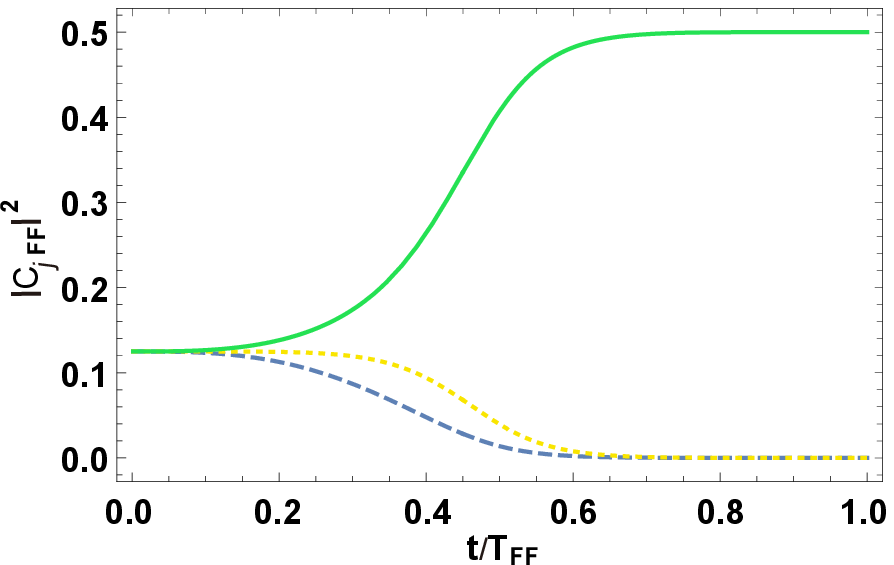}%
}
\caption{
The same time dependence as in Fig. \ref{fig-triangl}, but  in the case of the open linear 3 spin chain. 
(a) All eight eigenvalues;
(b) Regularization terms $v(t)\tilde{W}_{1}$ (dashed line) and $v(t)\tilde{W}_{2}$ (dotted line) ;
(c) Probability amplitudes for the solution $\Psi_{FF}(t)$ of TDSE,  $|C_3^{FF}|^2=|C_6^{FF}|^2$ (solid line), $ |C_1^{FF}|^2=|C_8^{FF}|^2$ (dashed line) and $|C_2^{FF}|^2=|C_4^{FF}|^2=|C_5^{FF}|^2=|C_7^{FF}|^2$ (dotted line).} 
\label{fig-linear3}
\end{figure}

In a similar way we can obtain the regularization term and fast-forward Hamiltonian in the case of open linear 3 spin chain.
In this case  the eigenvalue for the ground state is $E_0 = -\frac{1}{6}\big(B_x+(\beta+\bar{\beta}) -\sqrt{3}i(\beta-\bar{\beta})\big)$, where
$\beta=\big(18J^2B_x-8B_x^3+6Ji\sqrt{48J^4+39B_x^2J^2+24B_x^4} \big)^{1/3}$.
We have confirmed in Fig. \ref{fig-linear3} (a) that all eight eigenvalues show no mutual energy crossing in the fast-forward time range where we choose $J(R(\Lambda(t)))\equiv R(\Lambda(t))$
and $B_x(R(\Lambda(t)))\equiv B_0 - R(\Lambda(t))$ with $R(\Lambda(t))$ defined in Eq.(\ref{lambda2}).

The components of the eigenvector for the ground state are  :\\
$ C_1 =C_8= V_1 \zeta$, $C_2 =C_4 =C_5 = C_7 = V_2 \zeta$, $C_3=C_6= V_3 \zeta $, 
where $V_1 = {\frac {3{B_x}^{2}-8JB_x-4B_x E_0 -4E_0^2-8 E_0 J}{4JB_x}}$, 
$V_2 =-\frac{1}{2}V_1 - \frac{2J+E_0}{B_x}$,  $V_3=1$, and 
$\zeta=\frac{1}{\sqrt{2V_1^2+4V_2^2+2}}$. 

Here we see the symmetry: $C_1=C_8$, $C_2 =C_4 =C_5 = C_7$ and $C_3=C_6$. 
From $R$-derivative of the normalization ($\sum_{j=1}^8C_j^2=2C_1^2+4C_2^2+2C_3^2 = 1$), we see
\begin{equation}\label{NA-line3}
C_1 \frac{\partial C_1}{\partial R} + 2 C_2\frac{\partial C_2}{\partial R}+ C_3 \frac{\partial C_3}{\partial R} = 0,
\end{equation}
and then the adiabatic phase $\xi = 0$. 

The regularization Hamiltonian for the linear 3 spin system can also be available without using the 3-body interaction.  
Because of the geometric symmetry seen in Fig. \ref{3spinmodel}(b),
$\mathcal{\tilde{H}}$ is then characterized by two independent pairwise interactions:
$\tilde{W}_{1} \equiv \tilde{W}_{12}^{yz}=\tilde{W}_{23}^{yz}$ and $\tilde{W}_{2} \equiv \tilde{W}_{31}^{yz}$.
$\tilde{W}_{1}$ and $\tilde{W}_{2}$ correspond to the nearest-neighboring (N.N.) and 2nd N.N. interactions, respectively.
With use of the spin configuration bases, the matrix form for $\mathcal{\tilde{H}}$ in Eq. (\ref{cdterm}) is given by
\begin{widetext}
\begin{equation}\label{cdterm3spin}
\mathcal{\tilde{H}} =i \begin{pmatrix} 
0 &  -\tilde{W}_1-\tilde{W}_2& -2\tilde{W}_1 & -\tilde{W}_1-\tilde{W}_2 & 0 & 0 & 0 & 0  \\
\tilde{W}_1+\tilde{W}_2& 0 & 0& 0& 0 & -\tilde{W}_1+\tilde{W}_2 & 0 & 0 \\
2\tilde{W}_1 & 0 &  0 & 0 & \tilde{W}_1-\tilde{W}_2 & 0 &\tilde{W}_1-\tilde{W}_2& 0  \\
\tilde{W}_1+\tilde{W}_2& 0 & 0& 0& 0 & -\tilde{W}_1+\tilde{W}_2 & 0 & 0 \\
0 & 0 &  -\tilde{W}_1+\tilde{W}_2 & 0 &  0 & 0& 0 & \tilde{W}_1+\tilde{W}_2\\
0 &  \tilde{W}_1-\tilde{W}_2 & 0 &  \tilde{W}_1-\tilde{W}_2 & 0 & 0 & 0 & 2\tilde{W}_1\\
0 & 0 &  -\tilde{W}_1+\tilde{W}_2 & 0 &  0 & 0& 0 & \tilde{W}_1+\tilde{W}_2\\
0 & 0 & 0 & 0 & -\tilde{W}_1-\tilde{W}_2 & -2\tilde{W}_1& -\tilde{W}_1-\tilde{W}_2 &   0
\end{pmatrix}.
\end{equation}
\end{widetext}

Due to the symmetry of $\{C_j\}$, the number of independent equations in Eq. (\ref{sum2}) are three: 
\begin{eqnarray} \label{alg-line3}
-2(\tilde{W}_{1}+\tilde{W}_{2})C_2-2 \tilde{W}_{1}C_3&=& \hbar \frac{\partial C_1}{\partial R}
\nonumber\\
(\tilde{W}_{1}+\tilde{W}_{2})C_1+( -\tilde{W}_{1}+\tilde{W}_{2})C_3&=& \hbar \frac{\partial C_2}{\partial R}
\nonumber\\
2\tilde{W}_{1}C_1+2(\tilde{W}_{1}-\tilde{W}_{2})C_2&=& \hbar \frac{\partial C_3}{\partial R}.
\nonumber\\
\end{eqnarray}
By using Eq.(\ref{NA-line3}),  the 3rd line (for example) of the above equation proves trivial. Then Eq.(\ref{alg-line3}), whose coefficient matrix has the rank 2,
gives the solution:
\begin{eqnarray}\label{linear3}
\tilde{W}_{1}
&=&-\frac{\hbar}{2}\frac{\partial (C_1-C_3)}{\partial R}/(C_1+2C_2+C_3),\nonumber\\
\tilde{W}_{2}
&=&-\frac{\hbar}{2}\frac{\partial (C_1-2C_2+C_3)}{\partial R}/(C_1+2C_2+C_3).\nonumber\\
\end{eqnarray}
Including the regularization terms followed by rescaling of time, the fast forward Hamiltonian are written as 
\begin{equation}\label{HFF1}
H_{FF}= H_0(R(\Lambda(t)))+v(t) \tilde{\mathcal{H}} (R(\Lambda(t)))
\end{equation}
with $H_0$ = $J(R(\Lambda(t))) (\sigma_1^z \sigma_2^z+ \sigma_2^z \sigma_3^z)- \frac{1}{2}(\sigma_1^x+ \sigma_2^x+\sigma_3^x)B_x(R(\Lambda(t)))$, 
and $v\mathcal{\tilde{H}}$ = $v(t) \tilde{W}_{1}(R(\Lambda(t)))\big[ (\sigma_1^y \sigma_2^z+ \sigma_1^z \sigma_2^y)+(\sigma_2^y \sigma_3^z+ \sigma_2^z \sigma_3^y)\big]+v(t)\tilde{W}_{2}(R(\Lambda(t)))(\sigma_1^y \sigma_3^z+ \sigma_1^z \sigma_3^y)$.
The fast forward Hamiltonian guarantees the fast forward of the adiabatic dynamics of the ground state wave function.
Figures  \ref{fig-linear3}(b) and  \ref{fig-linear3}(c) show the time dependence of the  regularization terms  and that of the wave function, respectively. The wave function starts from the ground state  with $J=0$, i.e., $C_j = \frac{1}{2\sqrt{2}}$ for $j=1,\cdots, 8$.
As $J$  is increased from $0$ and $B_x$ is decreased, the system rapidly changes to the final state, i.e., a linear combination of reduced bases.  
In Fig. \ref{fig-linear3} (c) the solution $\Psi_{FF}(t)$ of TDSE in Eq.(\ref{TDSE})  has exactly reproduced the time-rescaled ground state wave function.

In case of $N=3$ spin systems, we have obtained the regularization terms and the fast-forward Hamiltonian without having recourse to the 3-body interaction.
Of course, we can see regularization terms which include the 3-body interaction: For a regular triangle we can have an extra solution consisting of only the 3-body interaction ($\tilde{Q}$), and for the open linear 3 spin system there can be solutions where $\tilde{Q}\neq 0$ and one of  $\tilde{W_1}$ and $\tilde{W_2}$ is non-vanishing. But these extra solutions are less interesting
from the viewpoint of searching for simpler controls.
In the case of $N=4$ spin systems in next Section, however, we cannot proceed without the 3-body interaction, although it will play only a subsidiary role.

\section{triangular pyramid, square, star graph and open linear 4 spin chain}\label{Four spins}

Now we shall investigate regular spin clusters with $N=4$ spins, namely, a triangular pyramid, square, star graph and open linear 4 spin chain in Fig. \ref{4spinmodel}.
Their original (reference) and regularization Hamiltonians are already given by Eq.(\ref{SIM}) and Eq.(\ref{cdterm}), respectively, where we put $N=4$.

By using the spin configuration bases, $\Ket{1}=\Ket{\uparrow\uparrow\uparrow\uparrow} $, $\Ket{2}=\Ket{\uparrow\uparrow\uparrow\downarrow} $, $\Ket{3}=\Ket{\uparrow\uparrow\downarrow\uparrow}$, $\Ket{4}=\Ket{\uparrow\downarrow\uparrow\uparrow}$,
$\Ket{5}=\Ket{\downarrow\uparrow\uparrow\uparrow}$
 $\Ket{6}=\Ket{\uparrow\uparrow\downarrow\downarrow}$, $\Ket{7}=\Ket{\uparrow\downarrow\downarrow\uparrow}$, $\Ket{8}=\Ket{\downarrow\downarrow\uparrow\uparrow}$, $\Ket{9}=\Ket{\downarrow\uparrow\uparrow\downarrow}$,
 $\Ket{10}=\Ket{\uparrow\downarrow\uparrow\downarrow}$,
 $\Ket{11}=\Ket{\downarrow\uparrow\downarrow\uparrow}$,
 $\Ket{12}=\Ket{\downarrow\downarrow\downarrow\uparrow}$,
 $\Ket{13}=\Ket{\downarrow\downarrow\uparrow\downarrow}$,
 $\Ket{14}=\Ket{\downarrow\uparrow\downarrow\downarrow}$,
  $\Ket{15}=\Ket{\uparrow\downarrow\downarrow\downarrow}$, and
   $\Ket{16}=\Ket{\downarrow\downarrow\downarrow\downarrow}$, 
the matrix form for original Hamiltonian $H_0$ in Eq.(\ref{SIM}) can be constructed.

\subsection{Triangular pyramid}
The eigenvalue of the ground state is
$E_0= \frac{1}{3} (-(\beta+\bar{\beta})+4J+\sqrt{3}i(\beta-\bar{\beta}) )$, where
$\beta=(35\,{J}^{3}-18{B_x}^{2}J+3i\sqrt {108{J}^{6}+309{B_x}^{2}{J}^{4}+3{B_x}^{4}{J}^{2}+3{B_x}^{6}})^{1/3}.$
For all regular clusters with $N=4$ spins in Fig. \ref{4spinmodel},  as is the case of the previous Section, we have numerically confirmed that there is no level crossing between the ground and 1st excited states in the fast-forward time range. So figures of 16 eigenvalues will be suppressed in this Section.

The components of the eigenvector of the ground state are:
$C_1=C_{16}=V_1\zeta$,
$C_2=C_3=C_4=C_5=C_{12}=C_{13}=C_{14}=C_{15}=V_2\zeta$, 
and
$C_6=C_7=C_8=C_9=C_{10}=C_{11}=V_6\zeta$. Here $\zeta=(2+8V_2^2+6V_6^2)^{-1/2}$, 
$V_1=1$, $V_2=\frac{(\beta+\bar{\beta})+14J-\sqrt{3}i(\beta-\bar{\beta})}{6B_x}$,
and $V_6=-{
\frac{2\left(\beta^2+\bar{\beta}^2\right) -10J \left(\beta+\bar{\beta}\right)
-\left(48 J^2+15B_x^2\right)
+i\sqrt{3}
(2\left(\beta^2-\bar{\beta}^2\right) +10J \left(\beta-\bar{\beta}\right))
}{27{B_x}^{2}}}$, where the equality $|\beta|^2=13J^2+3B_x^2$ is used.

From $R$-derivative of the normalization ($2C_1^2+8C_2^2+6C_6^2=1$), we see
\begin{equation}\label{NA-pyramid}
C_1 \frac{\partial C_1}{\partial R} + 4 C_2\frac{\partial C_2}{\partial R}+ 3C_6 \frac{\partial C_6}{\partial R} = 0 .
\end{equation}
If we suppress the 3-body interaction, the regularization Hamiltonian consists of only one pairwise interaction $\tilde{W} \equiv \tilde{W}_{ij}^{yz}$, due to the high symmetry of the triangular pyramid in Fig.\ref{4spinmodel} (a).
The corresponding matrix for the regularization term can be written as
\begin{widetext}
\begin{equation}\label{cdtermpyranid}
\mathcal{\tilde{H}} =i \left(
\begin{array}{cccccccccccccccc}
0 & -3\tilde{W} & -3\tilde{W} & -3\tilde{W} &  -3\tilde{W} & 0 & 0 & 0 & 0 & 0 & 0 & 0 & 0 & 0 & 0 & 0 \\
3\tilde{W} & 0 & 0 & 0 & 0 & -\tilde{W} & 0 & 0 & -\tilde{W} & -\tilde{W} & 0 & 0 & 0 & 0 & 0 & 0 \\
3\tilde{W} & 0 & 0 & 0 & 0 & -\tilde{W} & -\tilde{W} & 0 & 0 & 0 & -\tilde{W} & 0 & 0 & 0 & 0 & 0 \\
3\tilde{W} & 0 & 0 & 0 & 0 & 0 & -\tilde{W} & -\tilde{W} & 0 & -\tilde{W} & 0 & 0 & 0 & 0 & 0 & 0 \\
3\tilde{W} & 0 & 0 & 0 & 0 & 0 & 0 & -\tilde{W} & -\tilde{W} & 0 & -\tilde{W} & 0 & 0 & 0 & 0 & 0 \\
0 & \tilde{W} & \tilde{W} & 0 & 0 & 0 & 0 & 0 & 0 & 0 & 0 & 0 & 0 & \tilde{W} & \tilde{W} & 0 \\
0 & 0 & \tilde{W} & \tilde{W} & 0 & 0 & 0 & 0 & 0 & 0 & 0 & \tilde{W} & 0 & 0 & \tilde{W} & 0 \\
0 & 0 & 0 & \tilde{W} & \tilde{W} & 0 & 0 & 0 & 0 & 0 & 0 & \tilde{W} & \tilde{W} & 0 & 0 & 0 \\
0 & \tilde{W} & 0 & 0 & \tilde{W} & 0 & 0 & 0 & 0 & 0 & 0 & 0 & \tilde{W} & \tilde{W} & 0 & 0 \\
0 & \tilde{W} & 0 & \tilde{W} & 0 & 0 & 0 & 0 & 0 & 0 & 0 & 0 & \tilde{W} & 0 & \tilde{W} & 0 \\
0 & 0 & \tilde{W} & 0 & \tilde{W} & 0 & 0 & 0 & 0 & 0 & 0 & \tilde{W} & 0 & \tilde{W} & 0 & 0 \\
0 & 0 & 0 & 0 & 0 & 0 & -\tilde{W} & -\tilde{W} & 0 & 0 & -\tilde{W} & 0 & 0 & 0 & 0 & 3\tilde{W} \\
0 & 0 & 0 & 0 & 0 & 0 & 0 & -\tilde{W} & -\tilde{W} & -\tilde{W} & 0 & 0 & 0 & 0 & 0 & 3\tilde{W} \\
0 & 0 & 0 & 0 & 0 & -\tilde{W} & 0 & 0 & -\tilde{W} & 0 & -\tilde{W} & 0 & 0 & 0 & 0 & 3\tilde{W} \\
0 & 0 & 0 & 0 & 0 & -\tilde{W} & -\tilde{W} & 0 & 0 & -\tilde{W} & 0 & 0 & 0 & 0 & 0 & 3\tilde{W} \\
0 & 0 & 0 & 0 & 0 & 0 & 0 & 0 & 0 & 0 & 0 & -3\tilde{W} & -3\tilde{W} & -3\tilde{W} & -3\tilde{W} & 0 
\end{array}
\right) .
\end{equation}
\end{widetext}
Due to the symmetry of $\{C_j\}$, the number of independent equations in Eq.(\ref{sum2}) are three: 
\begin{eqnarray} \label{fuel}
-12 \tilde{W}C_2&=& \hbar \frac{\partial C_1}{\partial R}
\nonumber\\
3\tilde{W}C_1- 3\tilde{W}C_6&=& \hbar \frac{\partial C_2}{\partial R}
\nonumber\\
4\tilde{W}C_2&=& \hbar \frac{\partial C_6}{\partial R}.
\end{eqnarray}
While one of the above equations is trivial due to Eq.(\ref{NA-pyramid}),
we need one more unknown variable to make meaningful the algebraic equations in Eq. (\ref{fuel}).
Here we evaluate the contribution of the 3-body interaction.
The geometrical symmetry allows a universal 3-body interaction $\tilde{Q} \equiv \tilde{Q}_{ijk}^{xyz}$, independent of all possible 3-body configurations $(i,j,k)$. 
The inclusion of the 3-body interaction improves some matrix elements of  $\mathcal{\tilde{H}}$ in Eq.(\ref{cdtermpyranid})  as follows:
\begin{eqnarray}\label{3body-Q}
\mathcal{\tilde{H}}_{1,j}&=&\mathcal{\tilde{H}}_{16,j}=-4i\tilde{Q} \quad \rm{for} \quad \textit{j}=6,\cdots,11, \nonumber\\
\mathcal{\tilde{H}}_{i,1}&=&\mathcal{\tilde{H}}_{i,16}=4i\tilde{Q}  \quad \rm{for}  \quad \textit{i}=6, \cdots,11. \nonumber\\
\end{eqnarray}
After the above improvements, the algebraic equations in Eq.(\ref{fuel}) are revised as:
\begin{eqnarray} \label{fuel2}
-12 \tilde{W}C_2-24\tilde{Q}C_6&=& \hbar \frac{\partial C_1}{\partial R}
\nonumber\\
3\tilde{W}C_1- 3\tilde{W}C_6&=& \hbar \frac{\partial C_2}{\partial R}
\nonumber\\
8\tilde{Q}C_1+4\tilde{W}C_2&=& \hbar \frac{\partial C_6}{\partial R},
\end{eqnarray}
where one of the above lines is again trivial because of Eq. (\ref{NA-pyramid}). 
Equation (\ref{fuel2}), whose coefficient matrix has the rank 2, gives the solution:
\begin{eqnarray}\label{tr-py-Sol}
\tilde{W}&=&\frac{\hbar\partial_R C_2}{3(C_1-C_6)}, \nonumber\\
\tilde{Q}&=&\frac{\hbar\partial_R (C_1+3C_6)}{24(C_1-C_6)}.
\end{eqnarray}
The fast-forward Hamiltonian is given by
\begin{eqnarray}\label{HFF-tr-pyr}
H_{FF}&=& J(R(\Lambda(t))) \sum_{(i,j) \in N.N.}\sigma_i^z \sigma_j^z- \frac{1}{2}B_x(R(\Lambda(t)))\sum_{i=1}^{4}\sigma_i^x, \nonumber\\
&+&v(t) \tilde{\mathcal{H}} (R(\Lambda(t)))
\end{eqnarray}
with
\begin{eqnarray}\label{driv-tr-pyr}
v\mathcal{\tilde{H}}&=& \sum_{(i,j) \in all}v(t) \tilde{W}(R(\Lambda(t))) (\sigma_i^y \sigma_j^z + \sigma_i^z \sigma_j^y) \nonumber\\
&+& \sum_{(i,j,k) \in all}v(t)\tilde{Q}(R(\Lambda(t))) (\sigma_i^x \sigma_j^y + \sigma_i^y \sigma_j^x)\cdot \sigma_k^z.\nonumber\\
\end{eqnarray}
In the triangular pyramid, $\sum_{(i,j) \in N.N.}$ is equivalent to $\sum_{(i,j) \in all}$.
The fast forward Hamiltonian guarantees the fast forward of the adiabatic dynamics of the ground state wave function. 

\begin{figure}[H]
\subfloat[]{%
  \includegraphics[width=2in]{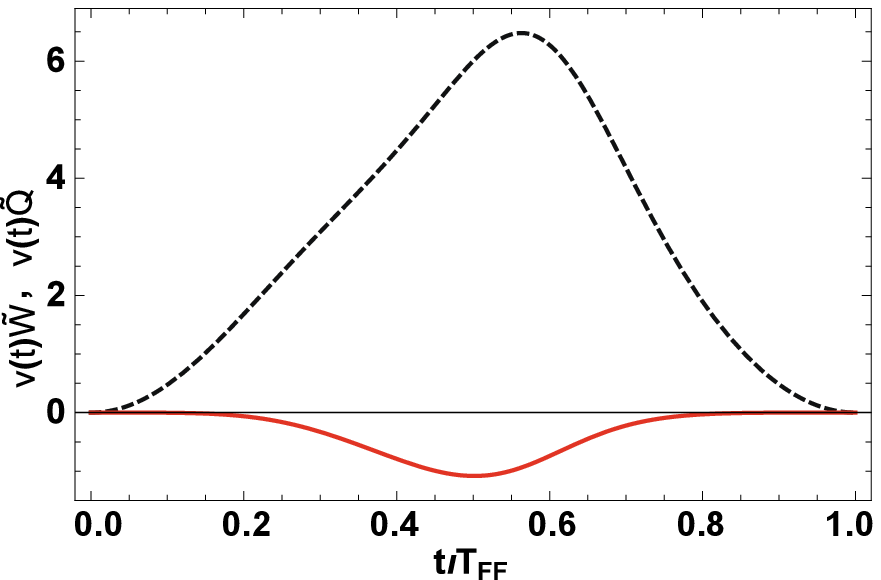}
}
\hfill
\subfloat[]{%
  \includegraphics[width=2in]{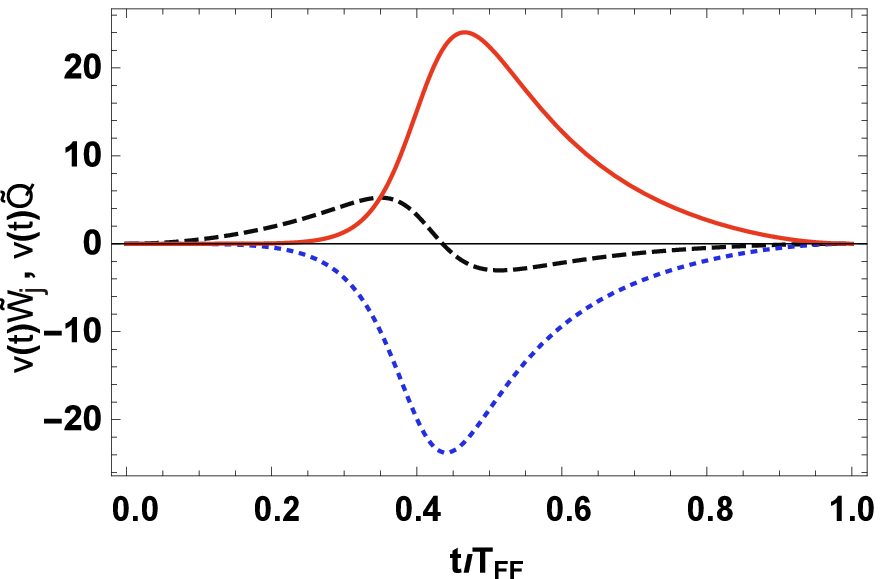}
}
\hfill
\subfloat[]{%
  \includegraphics[width=2in]{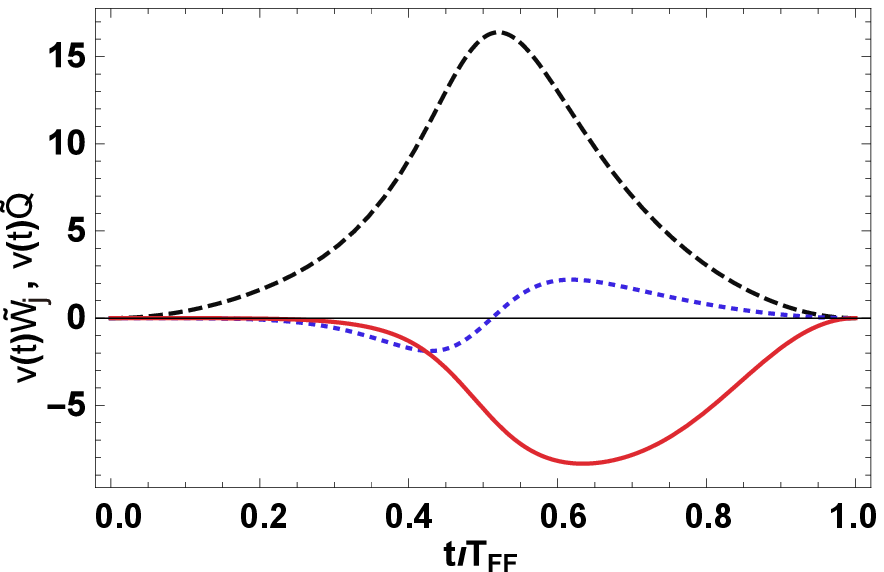}
}
\hfill
\subfloat[]{%
  \includegraphics[width=2in]{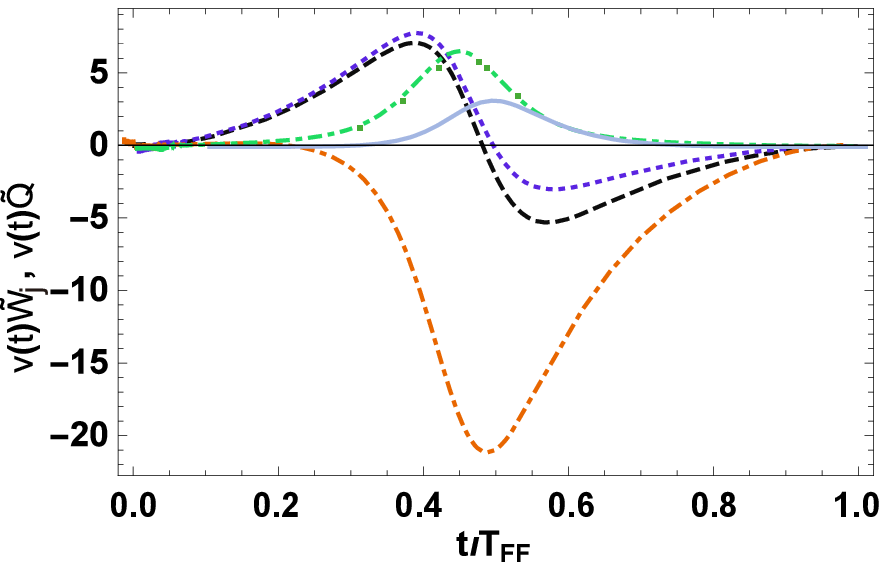}
}
\caption{The time dependence of regularization terms multiplied by $v(t)$  in the fast-forward time range where we choose $J=R(\Lambda(t))$ and $B_x = B_0-R(\Lambda(t))$ with 
$R(\Lambda(t))$ defined in Eq.(\ref{lambda2}). $B_0 =10$ and $\bar{v}= 100$.
 $T_{FF}=0.1$ and $R_0=0$. (a) Triangular pyramid. $v(t)\tilde{W}$ (dashed line) and $v(t)\tilde{Q}$ (solid line); (b) Square. $v(t)\tilde{W}_{1}$ (dashed line), $v(t)\tilde{W}_{2}$ (dotted line) and $v(t)\tilde{Q}$ (solid line) ; (c) Primary star graph. $v(t)\tilde{W}_{1}$ (dashed line), $v(t)\tilde{W}_{2}$ (dotted line) and $v(t)\tilde{Q}$ (solid line) ; (d) Open linear 4 spins. $v(t)\tilde{W}_{1}$ (dashed line), $v(t)\tilde{W}_{2}$ (dotted line), $v(t)\tilde{W}_{3}$ (dotted dashed line), $v(t)\tilde{W}_{4}$ (double-dotted dashed line) and $v(t)\tilde{Q}$ (solid line).} 
\label{fig-4spinreg}
\end{figure}

\begin{figure}[H]
\subfloat[]{%
  \includegraphics[width=2in]{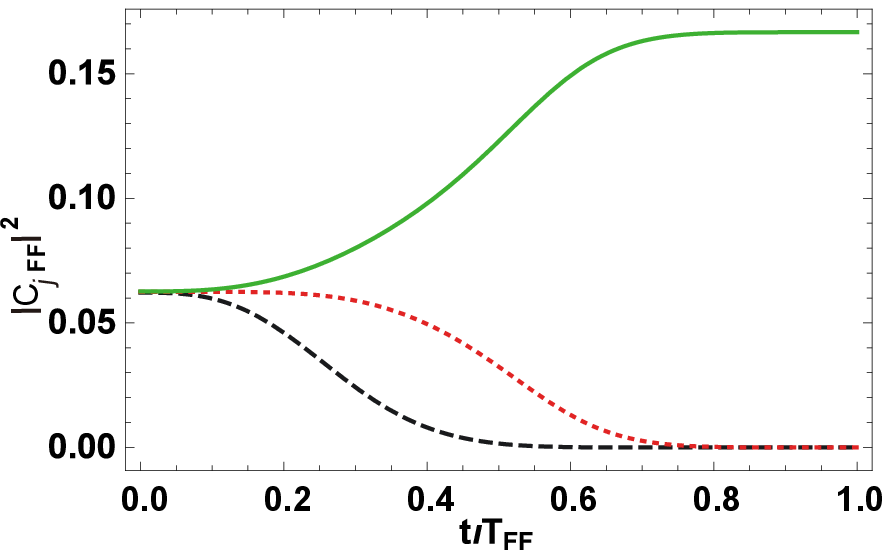}
}
\hfill
\subfloat[]{%
  \includegraphics[width=2in]{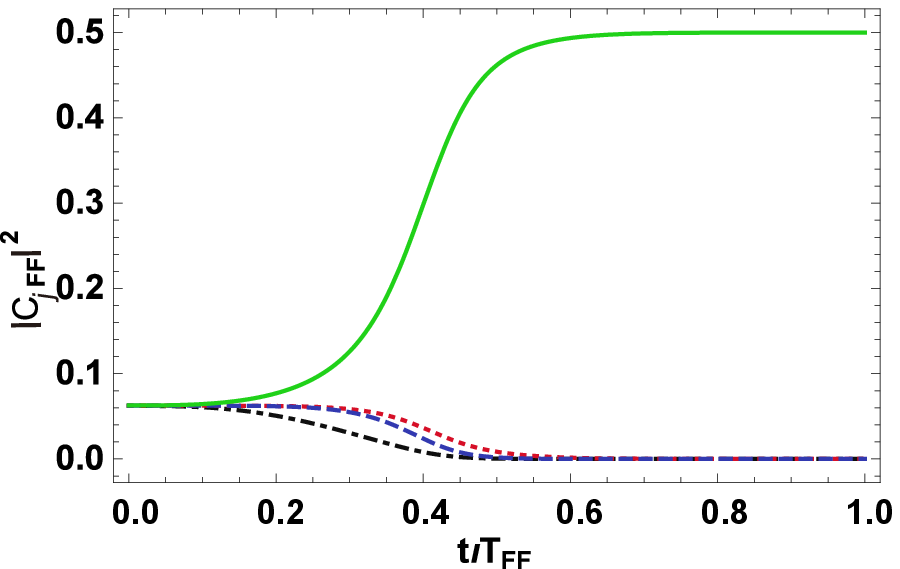}
}
\hfill
\subfloat[]{%
  \includegraphics[width=2in]{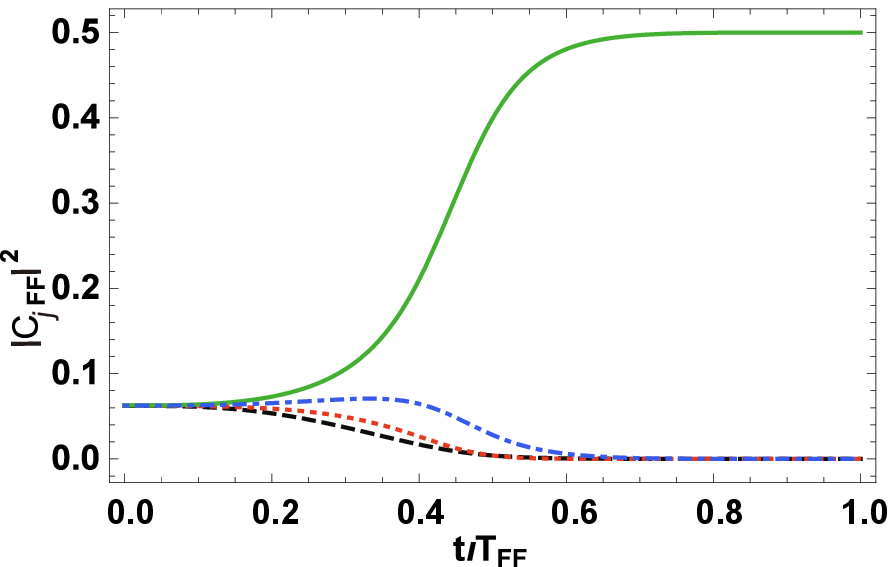}
}
\hfill
\subfloat[]{%
  \includegraphics[width=2in]{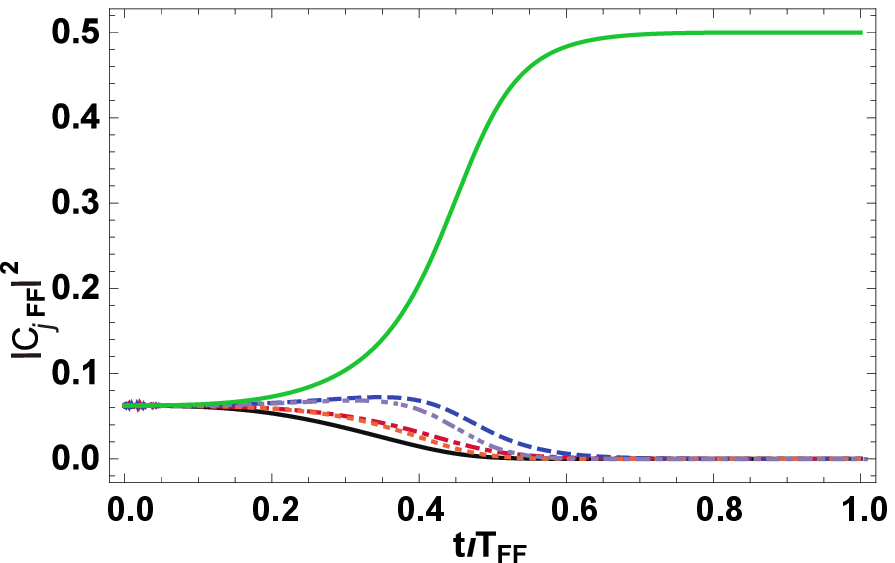}
}

\caption{Time dependence  of probability amplitudes $|C_j^{FF}|^2$ with $j=1\sim 16$ for the solution $\Psi_{FF}(t)$ of TDSE  in the fast-forward time range, where we choose $J=R(\Lambda(t))$ and $B_x = B_0-R(\Lambda(t))$ with $R(\Lambda(t))$ defined in Eq.(\ref{lambda2}). $B_0 =10$ and $\bar{v}= 100$.
 $T_{FF}=0.1$ and $R_0=0$. (a) Triangular pyramid. $j=1,16$ (dashed line), $ j=2 \sim 5$ and $12\sim15$ (dotted line), and $j=6 \sim 11$ (solid line); (b) Square. $j=1,16$ (dotted dashed line), $ j=2\sim5$ and $12\sim15$ (dotted line), $j=6\sim 9$ (dashed line), and $j=10,11$ (solid line); (c) Primary star graph. $j=1,16$ (dashed line), $ j=2,3,5,12,13,15$ (dotted line) , $j=4,14$ (solid line), and $j=6 \sim 11$ (dotted dashed line); (d) Open linear 4 spins.$j=1,16$ (lower solid line), $ j=2,5,12,15$ (dotted dashed line),  $j=3,4,13,14$ (dashed line), $j=6,8$ (dotted line), $j=7,9$ (double-dotted dashed line), and $j=10,11$(upper solid line).} 
\label{fig-4spinpro}
\end{figure}

Figures  \ref{fig-4spinreg} (a) and  \ref{fig-4spinpro} (a) show the time dependence of  regularization terms  and that of the wave function, respectively. The wave function starts from the ground state 
with $J=0$, i.e., $C_j=\frac{1}{4}$ for $j=1,\cdots, 16$.
In Fig. \ref{fig-4spinpro} (a) the solution $\Psi_{FF}(t)$ of TDSE in Eq.(\ref{TDSE})  has exactly reproduced the time-rescaled ground state wave function during the fast-forward time range $0 \leq t \leq T_{FF}$.

\subsection{Square}

The eigenvalue of the ground state is
$E_0=-\beta_1$, where
$\beta_1=\sqrt{8{J}^{2}+2{B_x}^{2}+2\beta_2}$ with $ \beta_2=\sqrt{16{J}^{4}+{B_x}^{4}}$.
The components of the eigenvector of the ground state are:
$C_1=C_{16}=V_1\zeta$,
$C_2=C_3=C_4=C_5=C_{12}=C_{13}=C_{14}=C_{15}=V_2\zeta$, 
$C_6=C_7=C_8=C_9=V_6\zeta$, and $C_{10}=C_{11}=V_{10}\zeta$ with $\zeta=(8+2V_1^2+4V_6^2+2V_{10}^2)^{-1/2}$. 
Here $V_1=\frac{(\beta_1-4J)(4J^2-B_x^2+\beta_2)}{8J^2B_x}$, $V_2=1$,
$V_6=\frac{(\beta_1^2-4\beta_2)\beta_1}{16J^2B_x}$, and $V_{10}=\frac{(\beta_1+4J)(4J^2-B_x^2+\beta_2)}{8J^2B_x}$.
From $R$-derivative of the normalization ($2C_1^2+8C_2^2+4C_6^2+2C_{10}^2=1$), we see
\begin{equation}\label{NA-square}
C_1 \frac{\partial C_1}{\partial R} + 4C_2\frac{\partial C_2}{\partial R}+ 2C_6 \frac{\partial C_6}{\partial R}+ C_{10} \frac{\partial C_{10}}{\partial R} = 0 
\end{equation}
The geometric symmetry of the square spin system in Fig.\ref{4spinmodel} (b) allows two candidates as regularization terms, which  are $\tilde{W}_{12}=\tilde{W}_{23}=\tilde{W}_{34}=\tilde{W}_{41}= \tilde{W}_1$ and $\tilde{W}_{31}=\tilde{W}_{42}= \tilde{W}_2$.
$\tilde{W}_1$ and $\tilde{W}_2$ correspond to N.N. and the second N.N. interactions, respectively. The regularization matrix $\mathcal{\tilde{H}}$ is given in Eq.(\ref{cdtermrec2}).
To add one more unknown variable, we include a contribution of the universal 3-body interaction $\tilde{Q}\equiv\tilde{Q}_{ijk}^{xyz}$. This inclusion requires the same improvement of some matrix elements of $\mathcal{\tilde{H}}$ as in Eq. (\ref{3body-Q}).

Due to the symmetry of $\{C_j\}$, the number of independent algebraic equations are four: 
\begin{eqnarray} \label{fuel-square}
(-8\tilde{W}_1- 4\tilde{W}_2)C_2-16\tilde{Q}C_6-8\tilde{Q}C_{10}&=& \hbar \frac{\partial C_1}{\partial R}
\nonumber\\
(2\tilde{W}_1+\tilde{W}_2)C_1+( -2\tilde{W}_2)C_6+ (-2\tilde{W}_1+\tilde{W}_2)C_{10}&=& \hbar \frac{\partial C_2}{\partial R}
\nonumber\\
8\tilde{Q}C_1+4\tilde{W}_2C_2&=& \hbar \frac{\partial C_6}{\partial R}.
\nonumber\\
8\tilde{Q}C_1+(8\tilde{W}_1-4 \tilde{W}_2)C_2&=& \hbar \frac{\partial C_{10}}{\partial R}.
\nonumber\\
\end{eqnarray}

Because of  Eq.(\ref{NA-square}), one of the above equations is trivial.  Ignoring the second line for example,
Eq.(\ref{fuel-square}), whose coefficient matrix has the rank 3, gives the solution:
\begin{eqnarray}\label{sq-solut}
\tilde{W}_1&=&-\frac{\hbar}{8(3C_1-C_{10}-2C_6)C_2}\times ( C_1\partial_RC_1-4C_2\partial_RC_2   \nonumber\\
&+&(C_1+C_{10})\partial_RC_6-(C_1-2C_6)\partial_RC_{10} ), \nonumber\\
\tilde{W}_2&=&-\hbar\frac{C_1\partial_RC_1-(C_1-2C_6-C_{10})\partial_RC_6+C_1\partial_RC_{10}}{4(3C_1-C_{10}-2C_6)C_2}, \nonumber\\
\tilde{Q}&=&\frac{\hbar\partial_R(C_1+2C_6+C_{10})}{8(3C_1-C_{10}-2C_6)C_2}.
\end{eqnarray}
The fast-forward Hamiltonian is given by Eq.(\ref{HFF-tr-pyr}), where $v(t)\tilde{\mathcal{H}} (R(\Lambda(t)))$ is now replaced by:
\begin{eqnarray}\label{driv-square}
v\mathcal{\tilde{H}}&=& \sum_{(i,j)=(1,2),(2,3),(3,4),(4,1)}v(t) \tilde{W}_1(R(\Lambda(t))) (\sigma_i^y \sigma_j^z + \sigma_i^z \sigma_j^y) \nonumber\\
&+&\sum_{(i,j)=(3,1),(4,2)}v(t) \tilde{W}_2(R(\Lambda(t))) (\sigma_i^y \sigma_j^z + \sigma_i^z \sigma_j^y) \nonumber\\
&+& \sum_{(i,j,k) \in all}v(t)\tilde{Q}(R(\Lambda(t))) (\sigma_i^x \sigma_j^y + \sigma_i^y \sigma_j^x)\cdot \sigma_k^z.\nonumber\\
\end{eqnarray}

Figures  \ref{fig-4spinreg} (b) and  \ref{fig-4spinpro} (b) show the time dependence of regularization terms  and that of wave function, respectively.  The wave function starts from the ground state 
with $J=0$, i.e., $C_j=\frac{1}{4}$ for $j=1,\cdots, 16$.
In Fig. \ref{fig-4spinpro} (b) the solution $\Psi_{FF}(t)$ of TDSE in Eq.(\ref{TDSE})  has exactly reproduced the time-rescaled ground state wave function.

\subsection{Primary star graph}
The eigenvalue of the ground state is
$E_0=-\beta$, where
$\beta=\sqrt{2{B_x}^{2}+5{J}^{2}+2\sqrt{{B_x}^{4}+{B_x}^{2}{J}^{2}+4{J}^{4}}}$.
The components of the eigenvector of the ground state are:
$C_1=C_{16}=V_1\zeta$,
$C_2=C_3=C_5=C_{12}=C_{13}=C_{15}=V_2\zeta$, 
$C_4=C_{14}=V_4\zeta$, and
$C_6=C_7=C_8=C_9=C_{10}=C_{11}=V_{6}\zeta$ with $\zeta=(6+2V_1^2+6V_2^2+2V_{4}^2)^{-1/2}$. 
Here $V_1=\frac{-J(7{B_x}^{2}+3J^2)
+\beta(4{B_x}^2+3\beta J-{\beta}^{2}+J^2)}{5J{B_x}^2}$, 
$V_2=\frac{-2J(9J^2-4{B_x}^{2})
+\beta(4{B_x}^2-2\beta J-{\beta}^{2}+21J^2)}{30J^2 B_x}$,
$V_4=\frac{-2J(J^2+4{B_x}^{2})
-\beta(4{B_x}^2-2\beta J-{\beta}^{2}+J^2)}{10J^2 B_x}$, and $V_6=1$. 
From $R$-derivative of the normalization  ($2C_1^2+6C_2^2+2C_4^2+ 6C_6^2=1$), we see
\begin{equation}\label{NA-pyramid2-star}
C_1 \frac{\partial C_1}{\partial R} + 3 C_2\frac{\partial C_2}{\partial R}+  C_{4} \frac{\partial C_4}{\partial R}+ 3C_6 \frac{\partial C_6}{\partial R} = 0. 
\end{equation}

The geometric symmetry of the primary star-graph spin system in Fig.\ref{4spinmodel}(c) allows two candidates as regularization terms, which are $\tilde{W}_{12}=\tilde{W}_{23}=\tilde{W}_{24}= \tilde{W}_1$ and $\tilde{W}_{14}=\tilde{W}_{13}=\tilde{W}_{34}= \tilde{W}_2$. 
$\tilde{W}_1$ and $\tilde{W}_2$ correspond to N.N. and the 2nd N.N. interactions, respectively. The matrix for regularization term $\mathcal{\tilde{H}}$ can be written in Eq.(\ref{cdtermrec1}).
To add one more unknown variable, we include a contribution of the universal 3-body interaction $\tilde{Q}\equiv\tilde{Q}_{ijk}^{xyz}$. This inclusion requires the same improvement of some matrix elements of $\mathcal{\tilde{H}}$ as in Eq. (\ref{3body-Q}).
One might have an idea to include two species of 3-body interactions with one among N.N.s and another among the 2nd N.N.s. But this idea results in incompatible
equations in Eq.(\ref{sum2}) and cannot be acceptable. Due to the symmetry of $\{C_j\}$, the number of independent equations are four: 
\begin{eqnarray} \label{fuel2-star}
(-6\tilde{W}_2-3\tilde{W}_1)C_2+( -3 \tilde{W}_1)C_4-24\tilde{Q}C_6&=& \hbar \frac{\partial C_1}{\partial R}
\nonumber\\
(2\tilde{W}_2+\tilde{W}_1)C_1+( -3\tilde{W}_1)C_6&=& \hbar \frac{\partial C_2}{\partial R}
\nonumber\\
(3\tilde{W}_1) C_1+(3\tilde{W}_1-6 \tilde{W}_2)C_6&=& \hbar \frac{\partial C_4}{\partial R}.
\nonumber\\
8\tilde{Q}C_1+(3\tilde{W}_1)C_2+(-\tilde{W}_1+2\tilde{W}_2)C_4&=& \hbar \frac{\partial C_{6}}{\partial R}.
\nonumber\\
\end{eqnarray}

Because of Eq.(\ref{NA-pyramid2-star}), one of the above 4 equations becomes trivial. Ignoring the first line for example,  Eq.(\ref{fuel2-star}), whose coefficient matrix has the rank 3, gives the solution: 
\begin{eqnarray}\label{star-solut}
\tilde{W}_1&=&\hbar\frac{C_1\partial_RC_4+3C_6\partial_RC_2}{3(C_1-C_{6})(C_1+3C_6)}, \nonumber\\
\tilde{W}_2&=&\hbar\frac{3(C_1+C_6)\partial_RC_2-(C_1-3C_6)\partial_RC_{4}}{6(C_1-C_{6})(C_1+3C_6)}, \nonumber\\
\tilde{Q}&=&\frac{\hbar}{24C_1(C_1-C_{6})(C_1+3C_6)} \nonumber\\
& \times & \bigl( 3(C_1^2+2C_1C_6-3C_6^2)\partial_RC_6 \nonumber\\
&-&3(3C_2C_6+C_1C_4)\partial_RC_2 \nonumber\\
& -&(3C_1C_2-2C_1C_4+3C_4C_6)\partial_RC_4\bigr).\nonumber\\
\end{eqnarray}
The fast-forward Hamiltonian is given by Eq.(\ref{HFF-tr-pyr}), where $v(t)\tilde{\mathcal{H}} (R(\Lambda(t)))$ is replaced by:
\begin{eqnarray}\label{driv-star}
v\mathcal{\tilde{H}}&=& 
\sum_{(i,j)=(1,2),(2,3),(2,4)}v(t) \tilde{W}_1(R(\Lambda(t))) (\sigma_i^y \sigma_j^z + \sigma_i^z \sigma_j^y) \nonumber\\
&+&\sum_{(i,j)=(1,4),(1,3),(3,4)}v(t) \tilde{W}_2(R(\Lambda(t))) (\sigma_i^y \sigma_j^z + \sigma_i^z \sigma_j^y) \nonumber\\
&+& \sum_{(i,j,k) \in all}v(t)\tilde{Q}(R(\Lambda(t))) (\sigma_i^x \sigma_j^y + \sigma_i^y \sigma_j^x)\cdot \sigma_k^z.\nonumber\\
\end{eqnarray}

Figures  \ref{fig-4spinreg} (c) and  \ref{fig-4spinpro} (c) show the time dependence of  regularization terms and that of wave function, respectively. The wave function starts from the ground state 
with $J=0$, i.e., $C_j=\frac{1}{4}$ for $j=1,\cdots, 16$.
In Fig. \ref{fig-4spinpro} (c) the solution $\Psi_{FF}(t)$ of TDSE in Eq.(\ref{TDSE})  has exactly reproduced the time-rescaled ground state wave function.

\subsection{Open linear 4 spin chain}

The eigenvalue of the ground state is
$E_0=-\frac{\beta_2}{\sqrt{3}}$, where
$\beta_2=\sqrt{2(\beta_1+\bar{\beta_1})+11J^2+4B_x^2}$ with 
$\beta_1=\bigl(64{J}^{6}+15{J}^{4}{B_x}^{2}+21{B_x}^{4}{J}^{2}+8{B_x}^{6}+3\sqrt {3}J^2B_x i\sqrt {128{J}^{6}+93{J}^{4}{B_x}^{2}+51{B_x}^{4}{J}^{2}+25{B_x}^{6} }
\bigr)^{1/3}$  
and $|\beta_1|^2=
4{B_x}^{4}+7{B_x}^{2}J^2+16J^4$.
The components of the eigenvector of the ground state are:
$C_1=C_{16}=V_1\zeta$,
$C_2=C_5=C_{12}=C_{15}=V_2\zeta$, 
$C_3=C_{4}=C_{13}=C_{14}=V_3\zeta$, 
$C_6=C_8=V_{6}\zeta$,  $C_7=C_9=V_{7}\zeta$, and  $C_{10}=C_{11}=V_{10}\zeta$ with $\zeta=(2+4V_2^2+4V_3^2+2V_6^2+2V_7^2+2V_{10}^2)^{-1/2}$. 
Here $V_1=1$, 
$V_2=-\frac {
\sqrt{3}{J}^{2}\beta_2\left(180{B_x}^{2} 
+144{J}^{2}\right)
-\sqrt{3}\beta_2^3 \left(12{B_x}^{2}
+33{J}^{2}\right)
+\sqrt{3} \beta_2^5
-162{J}^{5}}{ 162 {J}^{4}B_x}$,
$V_3=\frac {
\sqrt {3} {J}^{2} \beta_2\left(180{B_x}^{2} +198J^2\right) 
- \sqrt {3} \beta_2^3\left( 12 {B_x}^{2}  
+33{J}^{2} \right)
+\sqrt{3} \beta_2^5
+324{J}^{5}}{ 162 {J}^{4}B_x}$,
\begin{widetext}
$V_6=\frac {
-\sqrt {3} {J}^{2}\beta_2
(144 {B_x}^{2}+81{J}^{2})
-J\beta_2^2(36{B_x}^{2}+90 {J}^{2})
+\sqrt{3} \beta_2^3
(12{B_x}^{2}+30{J}^{2})
+3J \beta_2^4 
-\sqrt {3}
\beta_2^5+243{J}^{5}+ 648{B_x}^{2}{J}^{3}
}{216 {B_x}^{2}{J}^{3}}$, \\
$V_7=
-\frac {
\sqrt{3} {J}^{2}\beta_2 \left(
144 {B_x}^{2}
+81{J}^{2} \right) 
-J \beta_2^2 \left(36{B_x}^{2}
+90{J}^{2}\right)
-\sqrt{3}\beta_2^3\left(12{B_x}^{2} 
+30J^2\right)
+3J \beta_2^4
+\sqrt{3} \beta_2^5
+243{J}^{5}+324{J}^{3} B_x^2 }{108{B_x}^{2}{J}^{3} }$,
and \\
$V_{10}= -\frac {
\sqrt{3} J^2 \beta_2\left(144{B_x}^{2}-9{J}^{2}\right)
+ J \beta_2^2\left(
108{B_x}^{2}+54 {J}^{2}\right)
+6\sqrt {3}\beta_2^3\left(2{B_x}^{2}+J^2 \right)
-9J \beta_2^4 
-\sqrt{3} \beta_2^5
+648{B_x}^{2}{J}^{3} 
-81{J}^{5} 
} {648{B_x}^{2}{J}^{3} }$.
\end{widetext}
Since $\beta_2$ is real, all components of the ground state are also real. 
From $R$-derivative of the normalization  ($2C_1^2+4C_2^2+4C_3^2+ 2C_6^2+2C_7^2+2C_{10}^2=1$), we see
\begin{widetext}
\begin{equation}\label{NA-pyramid3}
C_1 \frac{\partial C_1}{\partial R} + 2 C_2\frac{\partial C_2}{\partial R}+2 C_{3} \frac{\partial C_3}{\partial R}+ C_6 \frac{\partial C_6}{\partial R}+ C_7 \frac{\partial C_7}{\partial R}+ 
C_{10} \frac{\partial C_{10}}{\partial R} = 0 
\end{equation}
\end{widetext}
In case of the open linear 4 spin system in Fig.\ref{4spinmodel}(d), the symmetry consideration allows 4 regularization terms which consist of  $\tilde{W}_{12}=\tilde{W}_{34}=\tilde{W}_1, \tilde{W}_{23}=\tilde{W}_2$, $\tilde{W}_{13}=\tilde{W}_{24}=\tilde{W}_3$, and $\tilde{W}_{14}= \tilde{W}_4$.
The regularization Hamiltonian $\mathcal{\tilde{H}}$ is given in Eq.(\ref{cdtermrec}).
To add one more unknown variable, we include a contribution of the universal 3-body interaction $\tilde{Q}\equiv\tilde{Q}_{ijk}^{xyz}$. This inclusion requires the same improvement of some matrix elements of $\mathcal{\tilde{H}}$ as in Eq. (\ref{3body-Q}).
The idea to include plural species of 3-body interactions results in incompatible equations in Eq.(\ref{sum2}) and can not be employed.
Due to the symmetry of $\{C_j\}$, the number of independent equations are six: 
\begin{widetext}
\begin{eqnarray} \label{fuel4}
(-2\tilde{W}_1-2\tilde{W}_4-2\tilde{W}_3)C_2+( -2 \tilde{W}_2-2\tilde{W}_1-2\tilde{W}_3)C_3-8\tilde{Q}C_6-8\tilde{Q}C_7-8\tilde{Q}C_{10}&=& \hbar \frac{\partial C_1}{\partial R}
\nonumber\\
(\tilde{W}_1+\tilde{W}_4+\tilde{W}_3)C_1+(-\tilde{W}_2+\tilde{W}_1-\tilde{W}_3)C_6+(-\tilde{W}_1+\tilde{W}_4-\tilde{W}_3)C_7+(-\tilde{W}_1-\tilde{W}_2+\tilde{W}_3)C_{10}&=& \hbar \frac{\partial C_2}{\partial R}
\nonumber\\
(\tilde{W}_1+\tilde{W}_2+\tilde{W}_3)C_1+(\tilde{W}_1-\tilde{W}_4-\tilde{W}_3)C_6+(-\tilde{W}_1+\tilde{W}_2-\tilde{W}_3)C_7+(-\tilde{W}_1-\tilde{W}_4+\tilde{W}_3)C_{10}&=& \hbar \frac{\partial C_3}{\partial R}
\nonumber\\
8\tilde{Q}C_1+(2\tilde{W}_2-2\tilde{W}_1+2\tilde{W}_3)C_2+( -2 \tilde{W}_1+2\tilde{W}_4+2\tilde{W}_3)C_3&=& \hbar \frac{\partial C_6}{\partial R}
\nonumber\\
8\tilde{Q}C_1+(2\tilde{W}_1-2\tilde{W}_2+2\tilde{W}_3)C_3+(2 \tilde{W}_1-2\tilde{W}_4+2\tilde{W}_3)C_2&=& \hbar \frac{\partial C_7}{\partial R}
\nonumber\\
8\tilde{Q}C_1+(2\tilde{W}_1+2\tilde{W}_2-2\tilde{W}_3)C_2+(2 \tilde{W}_1+2\tilde{W}_4-2\tilde{W}_3)C_3&=& \hbar \frac{\partial C_{10}}{\partial R} .
\end{eqnarray}
\end{widetext}
The constraint in Eq.(\ref{NA-pyramid3}) renders one of the above 6 equations trivial, and Eq.(\ref{fuel4}), whose coefficient matrix
has the rank 5, gives the following solution:
\begin{eqnarray}
\tilde{W}_1&=&\hbar \frac{\partial_R (C_7 + C_{10})}{4(C_2+C_3)} +\kappa, \nonumber\\
\tilde{W}_2&=&-\gamma_1+\gamma_2, \nonumber\\
\tilde{W}_3&=&\hbar\frac{\partial_R (C_6 + C_7)}{4(C_2+C_3)}+\kappa,\nonumber\\
\tilde{W}_4&=&\gamma_1+\gamma_2, \nonumber\\
\tilde{Q}&=&\frac{\hbar}{8( 3C_1-C_6-C_7- C_{10}) (C_1+C_6+C_7+C_{10})}\nonumber\\
&\times& \bigl(4(C_2-C_3)\partial_R (C_2-C_3)\nonumber\\
&+&(C_1+C_6+C_7+C_{10})\partial_R (C_1 + C_6+ C_7 + C_{10})\bigr),\nonumber\\
\end{eqnarray}

with
\begin{eqnarray}
\kappa&\equiv&\frac{\hbar}{2(C_2+C_3) (-3C_1+C_6+C_7+C_{10})} \nonumber\\
&\times&
\bigl(C_1\partial_R C_1
+(C_2 - C_3)\partial_R(C_2-C_3) \nonumber\\
&+&C_1\partial_R(C_6+C_7+C_{10})\bigr), \nonumber\\
\gamma_1&\equiv&\hbar\frac{\partial_R (C_2-C_3)}{2(C_1+C_6+C_7+C_{10})}, \nonumber\\
\gamma_2&\equiv&\hbar\frac{\partial_R (C_6+C_{10})}{4(C_2+C_3)} +\kappa.\nonumber\\
\end{eqnarray}

The fast-forward Hamiltonian is given by Eq.(\ref{HFF-tr-pyr}), where $v(t)\tilde{\mathcal{H}} (R(\Lambda(t)))$ is replaced by:
\begin{eqnarray}\label{driv-star}
v\mathcal{\tilde{H}}&=& 
\sum_{(i,j)=(1,2),(3,4)}v(t) \tilde{W}_1(R(\Lambda(t))) (\sigma_i^y \sigma_j^z + \sigma_i^z \sigma_j^y) \nonumber\\
&+&\sum_{(i,j)=(2,3)}v(t) \tilde{W}_2(R(\Lambda(t))) (\sigma_i^y \sigma_j^z + \sigma_i^z \sigma_j^y) \nonumber\\
&+&\sum_{(i,j)=(1,3),(2,4)}v(t) \tilde{W}_3(R(\Lambda(t))) (\sigma_i^y \sigma_j^z + \sigma_i^z \sigma_j^y) \nonumber\\
&+&\sum_{(i,j)=(1,4)}v(t) \tilde{W}_4(R(\Lambda(t))) (\sigma_i^y \sigma_j^z + \sigma_i^z \sigma_j^y) \nonumber\\
&+& \sum_{(i,j,k) \in all}v(t)\tilde{Q}(R(\Lambda(t))) (\sigma_i^x \sigma_j^y + \sigma_i^y \sigma_j^x)\cdot \sigma_k^z.\nonumber\\
\end{eqnarray}
Figures  \ref{fig-4spinreg} (d) and  \ref{fig-4spinpro} (d) show the time dependence of regularization terms and that of wave function, respectively. The wave function starts from the ground state 
with $J=0$, i.e., $C_j=\frac{1}{4}$ for $j=1,\cdots, 16$.
In Fig. \ref{fig-4spinpro} (d)  the solution $\Psi_{FF}(t)$ of TDSE in Eq.(\ref{TDSE})  has reproduced the time-rescaled ground state wave function, which means the perfect fidelity during the fast-forward time range $0 \leq t \leq T_{FF}$.

In this Section, the number of independent equations to determine the pair-wise interactions ($\tilde{W}_i$) is varied depending on the symmetry of clusters. To make Eq.(\ref{sum2}) solvable, however, these equations always require only one extra unknown 3-body interaction, whose contribution to $\mathcal{\tilde{H}}$ is commonly given in Eq.(\ref{3body-Q}) for all spin clusters with $N=4$ spins. Therefore the 3-body interaction  ($\tilde{Q}$)  here  is geometry-independent and played a subsidiary role.

\section{Summary and discussions}\label{concl}
The fast forward 
is the quasi-adiabatic dynamics
guaranteed by regularization terms added to the reference Hamiltonian,  followed by a rescaling of time with use of a large scaling factor. Assuming the regularization terms consisting of
pair-wise and 3-body interactions, we applied the core formula in Eq.(\ref{sum2})  to regular spin clusters with various geometries, e.g., regular triangle and open linear chain for $N=3$ spin systems, and triangular pyramid, square, primary star graph and open linear  chain for $N=4$ spin systems. 
The geometry-induced symmetry greatly decreases the rank of coefficient matrix of the linear algebraic equation for regularization terms, namely, the rank is determined by the geometric symmetry of the regular spin cluster.
Choosing a transverse Ising Hamiltonian as a reference, we find: \\
(1) for $N=3$ spin clusters, the driving interaction consists of only the geometry-dependent pair-wise interactions and there is no need for the 3-body interaction.
The regular triangle and open linear 3 spins require, respectively, one and 2 species of the pair-wise driving interactions; 
(2) for $N=4$ spin clusters, the main part of the driving interaction again consists of pair-wise interactions. The triangular pyramid and open linear 4 spins require, respectively, one and 4 species of the pair-wise driving interactions. On the other hand, two species of the pair-wise driving interactions are necessary for the square and primary star graph.
For $N=4$ spin clusters, besides these geometry-dependent pair-wise interactions, we need a common geometry-independent 3-body interaction just to make the core equation in Eq.(\ref{sum2}) solvable. The 3-body interaction here plays a subsidiary role.  The geometric symmetry of regular spin clusters determines the number of independent species of pair-wise driving interactions, and the clusters with the highest symmetry have only one species of pair-wise driving interaction. Our fast-forward scheme provides a flexible method in designing  the practical driving interaction in accelerating the adiabatic quantum dynamics of structured regular spin clusters. The scheme may also be useful in our inventing a variational method for treating much bigger regular clusters.

\begin{acknowledgements}
We are grateful to S. Masuda for valuable discussions in the early stage of the present work. The work is supported by Hibah Disertasi Doktor Kemenristekdikti 2018.  The work of B.E.G is supported by PUPT Ristekdikti-ITB 2017-2018.
\end{acknowledgements}

\appendix  

\section{Regularization matrix $\mathcal{\tilde{H}}$ without contributions due to the 3-body interaction}\label{apdB}
\subsection{Square}
The matrix for regularization term can be written as
\begin{widetext}
\begin{equation}\label{cdtermrec2}
\mathcal{\tilde{H}} =i \left(
\begin{array}{cccccccccccccccc}
0 & -A_1 & -A_1 & -A_1 &  -A_1 & 0 & 0 & 0 & 0 & 0 & 0 & 0 & 0 & 0 & 0 & 0 \\
A_1 & 0 & 0 & 0 & 0 & -A_2 & 0 & 0 & -A_2 & -A_3 & 0 & 0 & 0 & 0 & 0 & 0 \\
A_1 & 0 & 0 & 0 & 0 & -A_2 & -A_2 & 0 & 0 & 0 & -A_3 & 0 & 0 & 0 & 0 & 0 \\
A_1 & 0 & 0 & 0 & 0 & 0 & -A_2 & -A_2 & 0 & -A_3 & 0 & 0 & 0 & 0 & 0 & 0 \\
A_1 & 0 & 0 & 0 & 0 & 0 & 0 & -A_2 & -A_2 & 0 & -A_3 & 0 & 0 & 0 & 0 & 0 \\
0 & A_2 & A_2 & 0 & 0 & 0 & 0 & 0 & 0 & 0 & 0 & 0 & 0 & A_2 & A_2 & 0 \\
0 & 0 & A_2 & A_2 & 0 & 0 & 0 & 0 & 0 & 0 & 0 & A_2 & 0 & 0 & A_2 & 0 \\
0 & 0 & 0 & A_2 & A_2 & 0 & 0 & 0 & 0 & 0 & 0 & A_2 & A_2 & 0 & 0 & 0 \\
0 & A_2 & 0 & 0 & A_2 & 0 & 0 & 0 & 0 & 0 & 0 & 0 & A_2 & A_2 & 0 & 0 \\
0 & A_3 & 0 & A_3 & 0 & 0 & 0 & 0 & 0 & 0 & 0 & 0 & A_3 & 0 & A_3 & 0 \\
0 & 0 & A_3 & 0 & A_3 & 0 & 0 & 0 & 0 & 0 & 0 &A_3 & 0 & A_3 & 0 & 0 \\
0 & 0 & 0 & 0 & 0 & 0 & -A_2 & -A_2 & 0 & 0 & -A_3 & 0 & 0 & 0 & 0 & A_1 \\
0 & 0 & 0 & 0 & 0 & 0 & 0 & -A_2 & -A_2 & -A_3 & 0 & 0 & 0 & 0 & 0 & A_1 \\
0 & 0 & 0 & 0 & 0 & -A_2 & 0 & 0 & -A_2 & 0 & -A_3 & 0 & 0 & 0 & 0 & A_1 \\
0 & 0 & 0 & 0 & 0 & -A_2 & -A_2 & 0 & 0 & -A_3 & 0 & 0 & 0 & 0 & 0 & A_1\\
0 & 0 & 0 & 0 & 0 & 0 & 0 & 0 & 0 & 0 & 0 & -A_1 & -A_1 & -A_1 & -A_1 & 0 
\end{array}
\right)
\end{equation}
\end{widetext}
where $A_1$ = $2\tilde{W}_1 + \tilde{W}_2$, $A_2$ = $\tilde{W}_2$, $A_3$ = $2\tilde{W}_1-\tilde{W}_2$.

\subsection{Primary star graph}
The matrix for regularization term can be written as
\begin{widetext}
\begin{equation}\label{cdtermrec1}
\mathcal{\tilde{H}} =i \left(
\begin{array}{cccccccccccccccc}
0 & -A_1 &-A_1 & -A_2 &  -A_1 & 0 & 0 & 0 & 0 & 0 & 0 & 0 & 0 & 0 & 0 & 0 \\
A_1 & 0 & 0 & 0 & 0 & -A_3 & 0 & 0 & -A_3 & -A_3 & 0 & 0 & 0 & 0 & 0 & 0 \\
A_1 & 0 & 0 & 0 & 0 & -A_3 & -A_3& 0 & 0 & 0 & -A_3 & 0 & 0 & 0 & 0 & 0 \\
A_2 & 0 & 0 & 0 & 0 & 0 & A_4 & A_4 & 0 & A_4 & 0 & 0 & 0 & 0 & 0 & 0 \\
A_1 & 0 & 0 & 0 & 0 & 0 & 0 & -A_3 & -A_3 & 0 & -A_3 & 0 & 0 & 0 & 0 & 0 \\
0 & A_3 & A_3 & 0 & 0 & 0 & 0 & 0 & 0 & 0 & 0 & 0 & 0 & -A_4 &  A_3 & 0 \\
0 & 0 & A_3 & -A_4 & 0 & 0 & 0 & 0 & 0 & 0 & 0 & A_3 & 0 & 0 & A_3 & 0 \\
0 & 0 & 0 & -A_4 & A_3 & 0 & 0 & 0 & 0 & 0 & 0 & A_3 & A_3 & 0 & 0 & 0 \\
0 & A_3 & 0 & 0 & A_3 & 0 & 0 & 0 & 0 & 0 & 0 & 0 &A_3 & -A_4 & 0 & 0 \\
0 & A_3 & 0 & -A_4 & 0 & 0 & 0 & 0 & 0 & 0 & 0 & 0 & A_3 & 0 & A_3 & 0 \\
0 & 0 & A_3& 0 & A_3& 0 & 0 & 0 & 0 & 0 & 0 & A_3 & 0 & -A_4 & 0 & 0 \\
0 & 0 & 0 & 0 & 0 & 0 & -A_3 & -A_3 & 0 & 0 & -A_3 & 0 & 0 & 0 & 0 & A_1 \\
0 & 0 & 0 & 0 & 0 & 0 & 0 & -A_3 & -A_3 & -A_3 & 0 & 0 & 0 & 0 & 0 & A_1 \\
0 & 0 & 0 & 0 & 0 & A_4 & 0 & 0 & A_4 & 0 & A_4 & 0 & 0 & 0 & 0 & A_2 \\
0 & 0 & 0 & 0 & 0 & -A_3 & -A_3 & 0 & 0 & -A_3 & 0 & 0 & 0 & 0 & 0 & A_1 \\
0 & 0 & 0 & 0 & 0 & 0 & 0 & 0 & 0 & 0 & 0 & -A_1 & -A_1& -A_2 & -A_1 & 0 
\end{array}
\right)
\end{equation}
\end{widetext}
where $A_1$ = $2\tilde{W}_2+\tilde{W}_1$, $A_2$ = $3\tilde{W}_1$, 
$A_3$ = $\tilde{W}_1$, $A_4$ = $\tilde{W}_1-2\tilde{W}_2$.

\subsection{Open linear 4 spins}
The matrix for regularization term can be written as
\begin{widetext}
\begin{equation}\label{cdtermrec}
\mathcal{\tilde{H}} =i \left(
\begin{array}{cccccccccccccccc}
0 & -A_1 & -A_2 & -A_2 &  -A_1 & 0 & 0 & 0 & 0 & 0 & 0 & 0 & 0 & 0 & 0 & 0 \\
A_1 & 0 & 0 & 0 & 0 & -A_3& 0 & 0 & -A_4 & -A_5 & 0 & 0 & 0 & 0 & 0 & 0 \\
A_2 & 0 & 0 & 0 & 0 & -A_6 & -A_7 & 0 & 0 & 0 & -A_8 & 0 & 0 & 0 & 0 & 0 \\
A_2 & 0 & 0 & 0 & 0 & 0 & -A_7 & -A_6 & 0 & -A_8 & 0 & 0 & 0 & 0 & 0 & 0 \\
A_1 & 0 & 0 & 0 & 0 & 0 & 0 & -A_3 & -A_4 & 0 & -A_5 & 0 & 0 & 0 & 0 & 0 \\
0 & A_3 & A_6 & 0 & 0 & 0 & 0 & 0 & 0 & 0 & 0 & 0 & 0 & A_6 & A_3 & 0 \\
0 & 0 & A_7 & A_7 & 0 & 0 & 0 & 0 & 0 & 0 & 0 & A_4 & 0 & 0 & A_4 & 0 \\
0 & 0 & 0 & A_6 & A_3 & 0 & 0 & 0 & 0 & 0 & 0 & A_3 & A_6 & 0 & 0 & 0 \\
0 & A_4 & 0 & 0 & A_4 & 0 & 0 & 0 & 0 & 0 & 0 & 0 & A_7 & A_7 & 0 & 0 \\
0 & A_5 & 0 & A_8 & 0 & 0 & 0 & 0 & 0 & 0 & 0 & 0 & A_8 & 0 & A_5 & 0 \\
0 & 0 & A_8 & 0 & A_5 & 0 & 0 & 0 & 0 & 0 & 0 & A_5 & 0 & A_8 & 0 & 0 \\
0 & 0 & 0 & 0 & 0 & 0 & -A_4 & -A_3 & 0 & 0 & -A_5& 0 & 0 & 0 & 0 & A_1 \\
0 & 0 & 0 & 0 & 0 & 0 & 0 & -A_6 & -A_7 & -A_8 & 0 & 0 & 0 & 0 & 0 & A_2 \\
0 & 0 & 0 & 0 & 0 & -A_6 & 0 & 0 & -A_7 & 0 & -A_8 & 0 & 0 & 0 & 0 & A_2 \\
0 & 0 & 0 & 0 & 0 & -A_3 & -A_4 & 0 & 0 & -A_5 & 0 & 0 & 0 & 0 & 0 & A_1 \\
0 & 0 & 0 & 0 & 0 & 0 & 0 & 0 & 0 & 0 & 0 & -A_1 & -A_2 & -A_2 & -A_1 & 0 
\end{array}
\right)
\end{equation}
\end{widetext}
where 
$A_1$ = $\tilde{W}_1+\tilde{W}_3+\tilde{W}_4$, 
$A_2$ = $\tilde{W}_1+\tilde{W}_2+\tilde{W}_3$,  
$A_3$ = $\tilde{W}_2-\tilde{W}_1+\tilde{W}_3$, 
$A_4$ = $\tilde{W}_1-\tilde{W}_4+\tilde{W}_3$, 
$A_5$ = $\tilde{W}_1+\tilde{W}_2-\tilde{W}_3$, 
$A_6$ = $-\tilde{W}_1+\tilde{W}_4+\tilde{W}_3$, 
$A_7$ = $\tilde{W}_1-\tilde{W}_2+\tilde{W}_3$, 
$A_8$ = $\tilde{W}_1+\tilde{W}_4-\tilde{W}_3$.

\end{document}